\documentclass[sn-mathphys-num]{sn-jnl}

\usepackage{graphicx}%
\usepackage{epstopdf}
\usepackage{multirow}%
\usepackage{amsmath,amssymb,amsfonts}%
\usepackage{amsthm}%
\usepackage{mathrsfs}%
\usepackage[title]{appendix}%
\usepackage{xcolor}%
\usepackage{textcomp}%
\usepackage{manyfoot}%
\usepackage{booktabs}%
\usepackage{algorithm}%
\usepackage{algorithmicx}%
\usepackage{algpseudocode}%
\usepackage{listings}%
\begin{document}

\title[Longitudinal ZIP imputation]{Estimation and imputation of missing data in longitudinal models with Zero-Inflated Poisson  response variable}


\author[1]{\fnm{D. S.} \sur{Martinez-Lobo}}\email{dsmartinez@unbosque.edu.co}

\author[2]{\fnm{Oscar Orlando} \sur{Melo Martinez}}\email{oomelom@unal.edu.co}

\author*[3]{\fnm{Nelson Alirio} \sur{Cruz Gutierrez}}\email{nelson-alirio.cruz@uib.cat}

\affil[1]{\orgdiv{Profesor Asociado, Universidad del Bosque}}

\affil[2]{\orgdiv{Profesor Asociado, Departamento de  Estadistica, 
             Facultad de Ciencias,
            Universidad Nacional de Colombia}}

\affil*[3]{\orgdiv{ Profesor Visitante, Universitat de les Illes Balears, Departament de Matemàtiques i Informàtica}}


\abstract{This research deals with the estimation and imputation of missing data in longitudinal models with a Poisson response variable inflated with zeros. A methodology is proposed that is based on the use of maximum likelihood, assuming that data {is missing at random and that there is a correlation between the response variables}. In each of the times, the expectation maximization (EM) algorithm is used: in step E, a weighted regression is carried out, conditioned on the previous times that are taken as covariates. In step M, the estimation and imputation of the missing data are performed. {The good performance of the methodology in different loss scenarios is demonstrated in a simulation study comparing the model only with complete data, and estimating missing data using the mode of the data of each individual. Furthermore, in a study related to the growth of corn, it is tested on real data to develop the algorithm in a practical scenario.}}


%%================================%%
%% Sample for structured abstract %%
%%================================%%
\maketitle
\section{Introduction}

In data analysis, it is common to have missing information due to problems that arise during the obtaining of the information, being it experimental, observational, or sampling. The loss of information can lead to several issues in terms of the inference obtained for the population of interest. Some of these problems include biased estimators of the parameters and low efficiency of the estimators \citep{fitzmaurice2008longitudinal}.

In longitudinal studies, it is very common to have incomplete information, which constitutes one of the greatest challenges for the analysis of this type of data \citep{fitzmaurice2008longitudinal}. The most common situations that lead to the loss of information in this kind of research problem are the permanent withdrawal of subjects from the research, temporary withdrawal of subjects from the study, who later return to participate in later phases, and subjects joining the study after it has already started. These scenarios often result in the loss of valuable data.
It is clear that having missing information has implications for data analysis, and also alters the way of estimation of the usual models. Moreover, having incomplete data does not imply that there is a significant reduction in the precision of the estimates if the missing information is treated correctly; otherwise, biases are introduced in the estimators of the fitted model (\cite{ayala2007estimacion}, \cite{twisk2013applied} and \cite{lukusa2016semiparametric}).

The precision in the estimation depends on the amount of missing data, as well as the pattern by which the loss of data or subjects is generated. However, in most cases, the pattern of missing information is not within the control of the investigation. Therefore, it is necessary to make assumptions about the pattern of missing information. Thus, the validity of the results depends on how true the assumption is about the pattern of missing data. It is noteworthy that this research focuses on cases where the missing information is found in the response variable and not in the covariates \citep{lukusa2016semiparametric}.

In the literature, there is a wide variety of research on missing information. The investigations of \cite{dempster1977maximum} that propose to complete the missing information via the EM algorithm are highlighted, \cite{fitzmaurice1994analysing} described a method for the analysis of incomplete data in longitudinal studies with a binary response, \cite{ayala2007estimacion} estimated data missing in binary and Poisson response models via EM algorithm, \cite{daniels2008missing} used Bayesian methodologies to estimate missing data in longitudinal studies with a normal response, \cite{fitzmaurice2008longitudinal} showed a summary of the different current methodologies for estimating missing data in longitudinal models, \cite{chan2011bayesian} studied a bivariate longitudinal response using Bayesian methodologies to estimate missing data, \cite{twisk2013applied} performed analysis using generalized estimating equations, and \cite{lukusa2016semiparametric} estimated a model weighted linear for an inflated zero Poisson variable with missing data on the covariates.

In addition to the problem of longitudinal missing data, in many situations, the response variable is a count with many zeros, which are present in different disciplines such as agriculture \citep{hall2010robust}, sociology \citep{famoye2006zero}, dentistry \citep{mwalili2008zero}, psychometrics \citep{karazsia2008regression}, traffic accidents \citep{sharma2013zero} and health sciences \citep{yao2013semiparametric}, among others.

Consequently, zero-inflated distributions are a very good fit alternative in different research, and longitudinal models with zero-inflated response variables with missing information are not immune to modeling possibilities. \cite{lukusa2016semiparametric} showed a theoretical review of the existing literature relating zero-inflated data and missing data with different proposed estimation methods. It should be noted that no methodologies were found for the estimation of missing data in longitudinal models that take into account a Poisson-type response variable.

Therefore, the present investigation proposes a methodology for the estimation and imputation of the missing information when the longitudinal variable response has a zero-inflated Poisson distribution. {This methodology has several theoretical and practical advantages among which stand out: it is not necessary to specify a correlation structure within the observations of each experimental unit because the methodology uses them implicitly each time, and it gives better results in the estimation of the parameters and allows a very accurate prediction of missing data that improves as the presence of zeros is high in the response variable.}
The types of missing data and the notation that will be used in the paper are first presented. Then, the proposed methodology is shown with the necessary steps for the estimation and imputation of the missing data for the zero-inflated Poisson response variable. {Next, a simulation study comparing the model with only complete data and estimating the missing data using the mode of each individual's data is carried out.} Subsequently, the results obtained from the application are presented and analyzed from data taken from \cite{costa2003modelos}, related to a study of corn. Finally, the conclusions of the most important results of the research carried out are presented.

\section{Missing longitudinal data}
There are $T$ repeated measurements in time for each of the $n$ individuals that are part of the study. The response vector of the $i$th ($i=1\ldots, n$) individual at time $t$th ($t=1, \ldots,T$) is given by $\pmb{y }_{i}=(y_{i1},\ldots,y_{iT})'$ of size $n \times 1$. Associated to $\pmb{y}_{i}$ is the vector $\pmb{x}_{i}'$ of size $1 \times p$ with $p$ covariate values. Since there is missing information in the subject's responses, the vector $\pmb{y}_{i}$ divides into $\pmb{y}_{i}=(\pmb{y}_{i(miss) },\pmb{y}_{i(obs)})$, where $\pmb{y}_{i(obs)}$ is the vector of observed responses of size $n_{i}\times 1$, with $n_{i}$ the number of data points observed in the $i$th individual and $\pmb{y}_{i(miss)}$ the vector of missing responses of size $(T-n_{i} ) \times $1.
It is defined as $\pmb{R}_{i}=(R_{i1},\ldots,R_{iT})'$, with $R_{it}=1$ if $y_{it}$ is a response observed and $R_{it}=0$ if $y_{it}$ is a missing data. Using this notation, the different patterns of missing data in the \citep{rubin1976inference} responses are described below:

\begin{enumerate}
    \item Missing completely at random: These occur when the data is lost for reasons unrelated to the response variable or the covariates measured; therefore, these are independent of both the observed and unobserved values of the response variable. That is:
\begin{equation}
P(\pmb{R}_{i}\vert \pmb{y}_{i(miss)},\pmb{y}_{i(obs)},\pmb{x}_{i \times p }')=P(\pmb{R}_{i})
\end{equation}
In the case that the missing data are independent of the values of $\pmb{y}_{i}$, but depend on $\pmb{x}_{i \times p}'$, it is obtained that:
\begin{align}
P(\pmb{R}_{i}\vert \pmb{y}_{i(miss)},\pmb{y}_{i(obs)},\pmb{x}_{i \times p }')=P(\pmb{R}_{i}\vert \pmb{x}_{i \times p}')
\end{align}

This implies that the missing data can be explained by the covariates found in the data using a suitable model. Under the previous scheme, the resulting inference must be made on the information of the complete data set \citep{daniels2008missing}. It should be noted that for longitudinal studies this structure is very rare, since losing information in this way is very unusual, although cases have been reported in survival studies \citep{fernandes2008missing}.
\item Missing not at random: This structure occurs in cases where the probability of missing data depends on the value of the missing response or other unobservable values, that is,
  \begin{align}
P(\pmb{R}_{i}\vert \pmb{y}_{i(miss)},\pmb{y}_{i(obs)},\pmb{x}_{i \times p }')=P(\pmb{R}_{i}\vert \pmb{y}_{i(miss)},\pmb{x}_{i \times p}')
\end{align}
This type of structure makes it impossible to treat lost information. 
\item Missing at random: This case occurs when the missing data are independent of the missing responses and are conditional on the observed responses and the model covariates. This implies that the missing data depends on the observed data \cite{ayala2007estimacion}. The above can be written as:
  \begin{align}
P(\pmb{R}_{i}\vert \pmb{y}_{i(miss)},\pmb{y}_{i(obs)},\pmb{x}_{i \times p }')=P(\pmb{R}_{i}\vert \pmb{y}_{i(obs)},\pmb{x}_{i \times p}')
\end{align}
{The previous structure implies that the observed data cannot be seen as a random sample of the complete data contrary to the first structure. This structure should be the default assumption for the analysis of missing information in longitudinal data unless there is a strong and compelling reason to support another assumption \citep{fitzmaurice2008longitudinal}.}
\end{enumerate}
{Therefore, according to the previous description, from now on it will be assumed that missing data depends on the observed responses and the model of the covariates, i.e., missing at random. It will also be assumed that the explanatory variables do not have missing information, that is, the missing information is only presented in $\pmb{y}_{i}$, a common situation in experimental designs.} In addition, due to the structure of the data to be worked on, variable $y_{it}$ will be assumed to follow a zero-inflated Poisson distribution \citep{mwalili2008zero}.
\section{EM algorithm with ZIP}
Let $\pmb{Y}_{t}=(Y_{1t},\ldots, Y_{it}, \ldots,Y_{nt})'$ be the vector of the response variable observed at time $t$th for all individuals, it is known that the values $Y_{it}$ and $Y_{i't}$ are independent for each $i\neq i'$ \citep{lambert1992zero} and are defined as:
\begin{equation*}
Y_{it}=\begin{cases}
     0 & \text{with probability}\hspace{0.3cm} \pi_{it}\\
      \sim Poisson(\lambda_{it}) &\text{with probability }\hspace{0.3cm} 1-\pi_{it},\\
    \end{cases}
\end{equation*}
where $0\leq \pi_{it} \leq 1$ and $\lambda_{it} >0$. Therefore, the probability distribution of the random variable $Y_{it}$ is:
\begin{equation*}
     P(Y_{it}=y_{it})=\begin{cases}
         \pi_{it}+(1-\pi_{it})\exp{(-\lambda_{it})} &\text{if}\hspace{0.3cm} y_{it}=0,\\
                 \dfrac{(1-\pi_{it})\lambda_{it}^{y_{it}}\exp{(-\lambda_{it})}}{y_{it}!} &\text{if}\hspace{0.3cm} y_{it}>0,\\
     \end{cases}
\end{equation*}
Note that $E(Y_{it})=(1-\pi_{it})\lambda_{it}$ and the variance $Var(Y_{it})=(1-\pi_{it})\lambda_ {it}(1+\pi_{it} \lambda_{it})$. {To model the relationship between covariates and the response variable}, the link functions $\ln$ for $\lambda_{it}$ and \emph{logit} for $\pi_{it}$ are used as follows \citep{ ridout1998models}:
\begin{align*}
\ln(\lambda_{it})&=\mathbf{x}_{it}'\pmb{\beta}\\
\ln \left(\dfrac{\pi_{it}}{1-\pi_{it}}\right)&=\mathbf{z}_{it}' \pmb{\gamma}
\end{align*}
where $\pmb{z}_{it}'$ is the vector of associated covariates for the zero-occurrence model with size $1\times p_1$, $\pmb{\gamma}$ is the associated parameters vector of size $p \times 1$, $\pmb{x}_{it}'$ is the covariates vector associated with the Poisson model of size $ 1 \times p_2$ and $\pmb{\beta}$ is the parameters vector associated with the covariates of $\pmb{x}_{it}$ \cite{fang2013zero}. Clearing in each of the equations, it is obtained that:
\begin{equation}\label{1}
     \begin{split}
        \pi_{it}&= \dfrac{\exp(\pmb{z}_{it}'\pmb{\gamma})}{1+\exp(\pmb{z}_{it}'\pmb{\gamma })} \\
     \lambda_{it} &= \exp(\pmb{x}_{it}'\pmb{\beta})
      \end{split}
\end{equation}
In this zero-inflated model,  the parameter $\pi_{it}$ can be related to $\lambda_{it}$. This research assumes that the models that generate the zeros and the Poisson model are independent; so, it is assumed that $\pi_{it}$ is not related to $\lambda_{it}$. Therefore, the likelihood function of the vector $\pmb{Y}_{t}=(Y_{1t},\ldots, Y_{it}, \ldots,Y_{nt})'$ is determined as:
\begin{align} 
               \ln  \ell_t &=\ln  \prod_{i=1}^{n} P(Y_{it}=y_{it}) \nonumber \\
  &= \sum_{y_{it}=0}  \ln \left[ \exp(\pmb{z}_{it}'\pmb{\gamma}) + \exp(-\exp{(\pmb{x}_{it}'\pmb{\beta}))} \right] -\sum_{i=1}^{n}\left\{ \ln[ 1+\exp(\pmb{z}_{it}'\pmb{\gamma})] \right\} \nonumber \\
& \quad \quad     +
\sum_{y_{it}> 0}  \left\{y_{it} \pmb{x}_{it}'\pmb{\beta} - \exp(\pmb{x}_{it}'\pmb{\beta}) - \ln(y_{it}!) \right\} \label{4}
\end{align}
Let $D_{it}$ be and indicator function defined as:
\begin{equation*}
       D_{it} =
\left \{
	\begin{array}{ll}
		1 &\text{if } Y_{it}= 0 \\
		0&\text{if } Y_{it} > 0 \
	\end{array}
\right.
\end{equation*}
{where $D_{it}$ indicates the presence of zeros in the response of the experimental unit $i$-th at time $t$, in case the response is greater than 0, it will be worth 0. This indicator will allow dividing the likelihood functions in a part associated with the responses with a count equal to 0 and those without. So,} the $\log$-$\text{likelihood}$ function {of the complete data} is defined as:
\begin{align}
\ln \ell_t& =   \sum_{i=1}^{n} \left\{
\left( \ln[1+\exp{(\pmb{z}_{it}'\pmb{\gamma})}] + D_{it}\ln[  \exp{(\pmb{z}_{it}'\pmb{\gamma})}+ \exp{( -\exp{( \pmb{x}_{it}'\pmb{\beta} )}  )} ] \right)  \right. \nonumber \\
& \quad \left. +(1-D_{it}) \left[  -\ln[1+\exp{(\pmb{z}_{it}'\pmb{\gamma})}] -\exp{(\pmb{x}_{it}'\pmb{\beta})}+y_{it}\pmb{x}_{it}'\pmb{\beta}-\ln[y_{it}!] \right]    \right\} \label{l8}
\end{align}
Therefore, after algebraic calculations, the partial derivative of \eqref{l8} with respect to $\pmb{\beta}$ is:
\begin{align}\label{5}
\dfrac{\partial \ln \ell_t}{\partial \pmb{\beta}'} =&\sum_{i=1}^{n} \pmb{x}_{it}' 
D_{it}  \left[
\dfrac{-\exp{(\pmb{x}_{it}'\pmb{\beta})}\exp{(-\exp{(\pmb{x}_{it}'\pmb{\beta})})}}{\exp{(\pmb{z}_{it}'\pmb{\gamma})}+\exp{(-\exp{(\pmb{x}_{it}'\pmb{\beta})})}}
\right] + \nonumber\\
& \sum_{i=1}^{n}\pmb{x}_{it}' (1-D_{it})\left[-\exp{(\pmb{x}_{it}'\pmb{\beta})} +y_{it}\right] 
\end{align}
and the second partial derivative of \eqref{5} with respect to $\pmb{\beta}$ is:
\begin{align}
  \dfrac{\partial^{2} \ln \ell_t }{\partial \pmb{\beta} \partial \pmb{\beta}'}
=&\sum_{i=1}^{n} \pmb{x}_{it}' \Bigg[ \Bigg.
-D_{it}\exp{(\pmb{x}_{it}'\pmb{\beta})}\exp{(-\exp{(\pmb{x}_{it}'\pmb{\beta})})} \times  \nonumber  \\
&\left(
\dfrac{
\exp{(\pmb{z}_{it}'\pmb{\gamma})}+ \exp{(-\exp{(\pmb{x}_{it}'\pmb{\beta})})}-\exp{(\pmb{z}_{it}'\pmb{\gamma})}\exp{(\pmb{x}_{it}'\pmb{\beta})}
}{[\exp{(\pmb{z}_{it}'\pmb{\gamma})} +\exp{(-\exp{(\pmb{x}_{it}'\pmb{\beta})})} ]^{2}}
\right)\nonumber\\
&
+  (1-D_{it})[-\exp{(\pmb{x}_{it}'\pmb{\beta})}]
\Bigg. \Bigg] \pmb{x}_{it}\label{6}
\end{align}
If $ w_{it}=\exp{(\pmb{z}_{it}'\pmb{\gamma})}$ is replaced in \eqref{5} and \eqref{6}, respectively, it is obtained:
\begin{align*}
\dfrac{\partial \ln \ell_t }{\partial \pmb{\beta}'}=&\sum_{i=1}^{n} \pmb{x}_{it}' \left[\dfrac{-D_{it}\lambda_{it} \exp{(-\lambda_{it})}  }{w_{it}+\exp{(-\lambda_{it})}} +
    (1-D_{it})(y_{it}-\lambda_{it})    \right] \\
\dfrac{\partial^{2} \ln \ell_t }{\partial \pmb{\beta} \partial \pmb{\beta}'}    =&\sum_{i=1}^{n} \pmb{x}_{it}'\left[ -D_{it}\lambda_{it}\exp{(-\lambda_{it})}\left(
\dfrac{w_{it}+\exp{(-\lambda_{it})}-w_{it}\lambda_{it}  }{[w_{it}+\exp{(-\lambda_{it})}]^{2}}
\right)\right]\pmb{x}_{it} +\\
&\sum_{i=1}^{n} \pmb{x}_{it}'\left[(1-D_{it})[-\lambda_{it}]
\right]\pmb{x}_{it}
    \end{align*}
On the other hand, the partial derivative in \eqref{4} with respect to $\pmb{\gamma}$ is:
\begin{align} \label{9}
\dfrac{\partial \ln \ell_t }{\partial \pmb{\gamma}'}&=\sum_{i=1}^{n}\pmb{z}_{it}'\left\{\dfrac{-w_{it}}{1+w_{it}}+D_{it}\dfrac{w_{it}}{w_{it}+\exp{(-\lambda_{it})}}  \right\}\nonumber \\
&=\sum_{i=1}^{n}\pmb{z}_{it}'\left\{-\pi_{it}+D_{it}\dfrac{w_{it}}{w_{it}+\exp{(-\lambda_{it})}}  \right\}
\end{align}
The second partial derivative of \eqref{9} with respect to $\pmb{\gamma}$ in terms of $w_{it}$ and $\lambda_{it}$ is:
\begin{align*}
\dfrac{\partial^{2} \ln \ell_t}{\partial \pmb{\gamma} \partial \pmb{\gamma}'}&=\sum_{i=1}^{n} \left\{ -\pmb{z}_{it}'\left[ \pi_{it}(1-\pi_{it}) \right]\pmb{z}_{it}+
D_{it}\pmb{z}_{it}'\left[\dfrac{w_{it}\exp{(-\lambda_{it})}}{[w_{it}+\exp{(-\lambda_{it})}]^{2}} \right]\pmb{z}_{it} \right\} \\
&=\sum_{i=1}^{n}  \pmb{z}_{it}' \left[-\pi_{it}(1-\pi_{it})+\dfrac{D_{it}w_{it}\exp{(-\lambda_{it})}}{[w_{it}+\exp{(-\lambda_{it})}]^{2}}  \right]\pmb{z}_{it}
\end{align*}
Finally, partially differentiating \eqref{9} with respect to $\pmb{\beta}$, it is obtained that:
\begin{align*}
\dfrac{\partial^{2} \ln \ell_t}{\partial \pmb{\beta} \partial \pmb{\gamma}'}&=-\sum_{i=1}^{n} D_{it} \pmb{z}_{it}'\left[  \dfrac{w_{it}\lambda_{it}\exp{(-\lambda_{it})}}{[w_{it}+\exp{(-\lambda_{it})}]^{2}} \right]\pmb{x}_{it}
\end{align*}
Similarly, partially differentiating \eqref{5} with respect to $\pmb{\gamma}$ gives:

\begin{align*}
\dfrac{\partial^{2} \ln \ell_t}{\partial \pmb{\gamma} \partial  \pmb{\beta}'}=-\sum_{i=1}^{n} \pmb{x}_{it}' D_{it} \left[
\dfrac{\lambda_{it} w_{it}  \exp{(-\lambda_{it})}}{[w_{it}+\exp{(-\lambda_{it})}]^{2}} \right]   \pmb{z}_{it}
\end{align*}
Therefore, the Fisher information matrix $\pmb{\Im}_t$ has the form:
\begin{equation*}
    \begin{split}
        \pmb{\Im}_t&= \begin{pmatrix}
 \Im_{11t} & \Im_{12t} \\
 \Im_{21t} & \Im_{22t} \\
\end{pmatrix}
    \end{split}
\end{equation*}
where
\begin{equation*}
    \begin{split}
\Im_{11t}&=\sum_{i=1}^{n} \pmb{z}_{it}'\left[
\pi_{it}(1-\pi_{it})-\dfrac{D_{it}w_{it}\exp{(-\lambda_{it})} }{[w_{it}+\exp{(-\lambda_{it})} ]^{2}}
\right]\pmb{z}_{it} \\
\Im_{22t}&=\sum_{i=1}^{n}  \pmb{x}_{it}'\left[ (1-D_{it})(\lambda_{it}) \right. \\
& \left.   +
\dfrac{D_{it}\lambda_{it}\exp{(-\lambda_{it})}(w_{it}+\exp{(-\lambda_{it})}-w_{it}\lambda_{it})}{[w_{it}+\exp{(-\lambda_{it})} ]^{2}}
\right]\pmb{x}_{it} \\
\Im_{12t}&=\sum_{i=1}^{n} \pmb{z}_{it}'\left[\dfrac{D_{it}w_{it}\lambda\exp{(-\lambda_{it})}}{[w_{it}+\exp{(-\lambda_{it})}]^{2}} \right]\pmb{x}_{it}  \\
\Im_{21t}&=\sum_{i=1}^{n} \pmb{x}_{it}'\left[\dfrac{D_{it}w_{it}\lambda_{it}\exp{(-\lambda_{it})}}{[w_{it}+\exp{(-\lambda_{it})}]^{2}} \right]\pmb{z}_{it}
    \end{split}
\end{equation*}

The parameters estimated for $\pmb{\beta}$ and $\pmb{\gamma}$ cannot be obtained directly due to the complexity of the function of $\log$-$\text{likelihood}$, and the Scoring Fisher algorithm is not an option due {a part of the likelihood function depends on unobservable random variables, which are all the observations of $y_{it}$ that are missing} \citep{hall2010robust}. Then, a solution to maximize the $\log$-$\text{likelihood}$ function is by using the $EM$ algorithm proposed by \cite{mouatassim2012poisson}.
Therefore, the following is a detailed description of the algorithm proposed for the estimation and imputation of missing data when the response variable follows a zero-inflated Poisson distribution.
\begin{enumerate}
     \item For the time $t=1$, two steps are carried out: i)  {when the observation $y_{it}$ is missing, all possible values that this missing observation could take are generated} and with these, the parameter vector is estimated using the $EM$ algorithm {with a weight assigned to each possible value $k$ and that will be constructed in this section and associated with the estimated probability of occurrence of that value.}, and ii) the imputation is carried out taking into account the weights estimated in the previous step, and the model parameters are estimated again. 
     \item With the complete values for $t=1$, the algorithm is repeated for time $t=2$. {In addition to using the covariates of the model to explain the response variable $t=2$, will also be used as an explanatory response variable at time $t=1$. Therefore, the correlation that exists between the observations of the experimental unit will be modeled by its associated regression parameter.}
     \item Continue repeating steps 1 and 2, for each time $t>2$ until all times will be considered. {As for each time $t>2$, there will be at least two previous times of each experimental unit that could be correlated with each other, a principal component analysis (PCA) will be used between them before being included in the estimation model. This step allows us to take into account the possible correlation within each experimental unit without having to model it explicitly.}
\end{enumerate}

\subsection{Time 1}
From the function of $\log$-$\text{likelihood}$ expressed in \eqref{l8}, setting $t=1$, it is obtained that:
\begin{align}\label{10}
 \ln \ell_t&= \sum_{i=1}^{n}\left\{ D_{i1}
\ln[\exp{(\pmb{z}_{i1}'\pmb{\gamma})}+\exp{(-\exp{(\pmb{x}_{i1}'\pmb{\beta})})} ]-\ln[1+\exp{(\pmb{z}_{i1}'\pmb{\gamma})}] \right\} \nonumber \\
& +\sum_{i=1}^{n}  (1-D_{i1})\left(y_{i1}\pmb{x}_{i1}'\pmb{\beta}-\exp{(\pmb{x}_{i1}'\pmb{\beta})}-\ln[y_{i1}!] \right)
\end{align}
where $y_{i1}$ is the response vector of size $n \times 1$, $\pmb{x}_{i1}$ is the covariates vector associated with the Poisson model, and $\pmb{z}_{ i1}$ is the covariates vector of the zero model for the $i$th observation at time 1 of size $1 \times p$, where $p$ is the number of covariates. The maximization of the function $\log$-$\text{likelihood}$ given in \eqref{10} is complicated by the term $\ln[\exp{(\pmb{z}_{i1}'\pmb{ \gamma})}+\exp{(-\exp{(\pmb{x}_{i1}'\pmb{\beta})})} ]$
since it involves the two parameters that are sought to be estimated. But if the indicator function is $D_{i1}=0$, then:
\begin{align*}
D_{i1}\ln[\exp{(\pmb{z}_{i1}'\pmb{\gamma})}+\exp{(-\exp{(\pmb{x}_{i1}' \pmb{\beta})})} ]=0
\end{align*}
That is, when $y_{i1}=0$ then $D_{i1}=1$. So, this part of the function will estimate only the part of the model that is equal to zero for $y_{i1}$, and therefore, the covariates associated with the Poisson model are not taken into account in the estimation\cite{lambert1992zero}. Thus, the term that involves the two parameters that are sought to be estimated is reduced to:
\begin{align*}
D_{i1}\ln[\exp{(\pmb{z}_{i1}'\pmb{\gamma})}]=D_{i1}\pmb{z}_{i1}'\pmb{\gamma }
\end{align*}
Then, the expression \eqref{10} of the function $\log$-$\text{likelihood}$ reduces to:
\begin{align}
    \ln \ell&=\sum_{i=1}^{n}\left\{ D_{i1}\pmb{z}_{i1}'\pmb{\gamma}-\ln[1+\exp{(\pmb{z}_{i1}'\pmb{\gamma})}] \right\}\nonumber \\
    &+
\sum_{i=1}^{n}
(1-D_{i1})\left(y_{i1}\pmb{x}_{i1}'\pmb{\beta}-\exp{(\pmb{x}_{i1}'\pmb{\beta})}-\ln[y_{i1}!] \right)\nonumber\\
&=\ell(D_{i1},\pmb{\gamma}, y_{i1}) +\ell(D_{i1},\pmb{\beta},y_{i1})\label{13}
\end{align}
where
\begin{equation}\label{14}
    \begin{split}
\ell(D_{i1},\pmb{\gamma}, y_{i1})&=\sum_{i=1}^{n}\left\{ D_{i1}\pmb{z_{i1}}'\pmb{\gamma}-\ln[1+\exp{(\pmb{z}_{i1}'\pmb{\gamma})}] \right\}\\
\ell(D_{i1},\pmb{\beta},y_{i1})&=\sum_{i=1}^{n} (1-D_{i1})\left(y_{i1}\pmb{x}_{i1}'\pmb{\beta}-\exp{(\pmb{x}_{i1}'\pmb{\beta})}-\ln[y_{i1}!] \right)
       \end{split}
\end{equation}
Now, given that $y_{i1}=(y_{i1(obs)}, y_{i1(miss)})$ is the vector that refers to the observed and missing values of the response variable, the functions $\ell(D_{i1},\pmb{\gamma}, y_{i1})$ and $\ell(D_{i1},\pmb{\beta}, y_{i1})$ can be maximized separately, using the $EM$ algorithm. This algorithm works iteratively, starting with an initial estimate of $D_{i1}^{(0)}$ given the conditional expectation under some initial parameters $(\pmb{\gamma}^{(0)}, \pmb{\beta}^{(0)})$. Then, with $D_{i1}^{(0)}$ estimated, the maximization for $\pmb{\beta}^{(1)}$ and $\pmb{\gamma}^{(1)}$ is obtained, and so on until the $m$th iteration meets the convergence criteria.

From $\ell(D_{i1},\pmb{\gamma}, y_{i1})$, the likelihood for the step $m>1$, the expectation is written as:
\begin{align}\label{49}
    Q_{i1}(D_{i1},\pmb{\gamma} \vert ,D_{i1}^{(m)},\pmb{\gamma}^{(m)} ) &=E[\ell(D_{i1},\pmb{\gamma}, y_{i1})\vert x_{i1},D_{i1}=D_{i1}^{(m)},\pmb{\gamma}=
    \pmb{\gamma}^{(m)}]\nonumber \\
     &=\sum_{i=1}^{n}  \ell(D_{i1},\pmb{\gamma}, y_{i1})w_{i1(k)}^{(m)}
\end{align}
where { $w_{i1(k)}^{(m)}=P(y_{i1(miss)}=k\vert \pmb{x}_{i1},D_{i1}^{(m)},\pmb{\gamma}^{(m)})$ is the probability that a data item $y_{i1(miss)}$ is equal to the value $k$ conditional to the values of the parameters of the model and the values of $\pmb{x}$}.
Analogously, from $\ell(D_{i1},\pmb{\beta}, y_{i1})$, it is obtained that:
\begin{equation}\label{52}
Q_{i1}(D_{i1},\pmb{\beta} \vert  D_{i1}^{(m)},\pmb{\beta}^{(m)} ) =
 \sum_{i=1}^{n} \ell(D_{i1},\pmb{\beta}, y_{i1})w_{i1(k)}^{(m)}
\end{equation}
The equations \eqref{49} and \eqref{52} correspond to the likelihoods for the parameters $\pmb{\gamma}$ and $\pmb{\beta}$ with the complete weighted data. For each missing value, values {of $k$} are taken that are generated taking into account the nature of the response variable. That is since the random variable is composed of a Poisson model and an excess zero model, $y_{i1(miss)}\in \{0,1,2,3,\ldots\}$. Since handling infinitely many values is impossible numerically, an upper bound is defined from the central limit theorem, that is, if $Y\sim P(\lambda) \approx \sim N(\lambda,\lambda)$, then approximately $99\%$ of the values of the distribution lie in the interval $[\mu-3\sigma, \mu+3\sigma]$. Therefore, an upper bound is defined as $E[y_{i1( obs)}]+3\sqrt{E[y_{i1(obs)}]}$. 

For example, if the random variable has an expectation of $4$, the upper bound is $4+3\sqrt{4}=10$; therefore, the values that $y_{i1(miss)}$ can take are $0,1,2,3,4,5,6,7,8,9$ and $10$. The weights for the Poisson Zero Inflation model are defined taking into account the proposal of \cite{ayala2007estimacion} as follows:
{In case the $y_{i1}$ is observed, only its real value will be used in the likelihood with a weight of 1.} {If $y_{i1}$ is missing, its first possible value is $k=0$, and therefore:}
\begin{align*}
w_{i1(0)}^{(m)}&=[y_{i1(miss)}=0\vert \pmb{x}_{i1(miss)}, \pmb{\beta}^{(m)},\pmb{\gamma}^{(m)}]\\
&=\pi_{i1}^{(m)}+(1-\pi_{i1}^{(m)})\exp{(-\lambda_{i1}^{(m)})}
\end{align*}
{where $\pmb{x}_{i1(miss)}$ refers to the observed values of $\pmb{x}_{i1}$ for the missing observation $y_{i1}$}. If $k=1$, it is defined that $y_{i1}=1$, then
\begin{align*}
w_{i1(1)}^{(m)}&=P[y_{i1(miss)}=1\vert \pmb{x}_{i1(miss)}, \pmb{\beta}^{(m)},\pmb{\gamma}^{(m)}]\\
&=\dfrac{(1-\pi_{i1}^{(m)})\lambda_{i1}^{(m)}\exp{(-\lambda_{i1}^{(m)})}}{1!}
\end{align*}
Similarly, for an arbitrary value $k>1$ , $y_{i1}=k$, then
\begin{align*}
w_{i1(k)}^{(m)}&=P[y_{i1(miss)}=k\vert \pmb{x}_{i1(miss)}, \pmb{\beta}^{(m)},\pmb{\gamma}^{(m)}]\\
&=\dfrac{(1-\pi_{i1}^{(m)})(\lambda_{i1}^{(m)})^{k}\exp{(-\lambda_{i1}^{(m)})}}{(k)!}
\end{align*}
{Finally, if the data $y_{i1}$ is observed, $w_{i1(k)}^{(m)}=1$ if $k=y_{i1}$ and $w_{i1(k)}^{(m)}=0$ if $k\neq y_{i1}$. Therefore, the weights for the model are:
\begin{align*}
 w_{i1(k)}^{(m)} &= \begin{cases}
1 &\text{if the data $y_{i1}$ is observed and $k=y_{i1}$}\\
0 &\text{if the data $y_{i1}$ is observed and $k\neq y_{i1}$}\\
\pi_{i1}^{(m)}+(1-\pi_{i1}^{(m)})\exp{(-\lambda_{i1}^{(m)})} &\text{if the data $y_{i1}$ is missing and} \  k= 0\\
\dfrac{(1-\pi_{i1}^{(m)})\exp{(-\lambda_{i1}^{(m)})} (\lambda_{i1}^{(m)})^{k}}{(k)!} &\text{if ithe data $y_{i1}$ is missing and} \  k> 0\\
\end{cases} \\
\end{align*}
}

Now, in the next section, the estimated values for $\pmb{\gamma}$ and $\pmb{\beta}$ will be obtained using the $EM$ algorithm for the data for the time $t=1$.
\subsubsection{Step 1}

\paragraph{Step E.} From the definition of the Poisson Zero Inflation model, it is obtained that:
\begin{align*}
     (Y_{i1}\vert D_{i1}=1)&=0\\
     (Y_{i1}\vert D_{i1}=0)& \sim P(\lambda_{i1})
\end{align*}
Then, the expected value of $D_{i1}$ under the initial values for $\pmb{\beta}^{(m)}$ and $\pmb{\gamma}^{(m)}$ is written as:
\begin{align*}
E[D_{i1}^{(m)}\vert y_{i1},\pmb{ \gamma}^{(m)}, \pmb{\beta}^{(m)}]&=P[D_ {i1}^{(m)}=1\vert y_{i1},\pmb{\gamma}^{(m)}, \pmb{\beta}^{(m)}]
\end{align*}
and using Bayes' theorem, it is obtained that:
{\footnotesize
\begin{align*}
&P\left[D_{i1}^{(m)}=1\vert y_{i1},\pmb{\gamma}^{(m)},\pmb{\beta}^{(m)} \right]=\\
&\dfrac{  P[D_{i1}^{(m)}=1] P[y_{i1}=0\vert D_{i1}^{(m)}=1,\pmb{\gamma}^{(m)}, \pmb{\beta}^{(m)}]}{
P[D_{i1}^{(m)}=1] P[y_{i1}=0\vert D_{i1}^{(m)}=1,\pmb{\gamma}^{(m)}, \pmb{\beta}^{(m)}]+
P[D_{i1}^{(m)}=0] P[y_{i1}> 0\vert D_{i1}^{(m)}=0,\pmb{\gamma}^{(m)},\pmb{ \beta}^{(m)}]
}
\end{align*}
}
\noindent By the definition of the indicator function it is obtained that $P[D_{i1}^{(m)}=1]=\pi_{i1}$, $P[D_{i1}^{(m)}=0 ]=1-\pi_{i1}$ and also $E(D_{i1}^{(m)})=P[D_{i1}^{(m)}=1]$.  Therefore,
\begin{align*}
P[D_{i1}^{(m)}=1\vert y_{i1},\pmb{\gamma}^{(m)}, \pmb{\beta}^{(m)}] &= \begin{cases}
\dfrac{1}{1+ \exp{( -\exp{(\pmb{x}_{i1}'\pmb{\beta}^{(m)})}-\pmb{z}_{i1}'\pmb{\gamma}^{(m)}   ) } }  &  \text{if }  y_{i1}=0 \\
0 &\text{if} \  y_{i1}> 0\\
\end{cases}\\
&=E[D_{i1}^{(m)}\vert y_{i1},\pmb{\gamma}^{(m)}, \pmb{\beta}^{(m)}]
\end{align*}
\paragraph{Step M.} Fit $\pmb{\gamma}^{(m)}$ by the maximization of $\ell(D_{i1}^{(m)},\pmb{\gamma}^{ (m)},\pmb{y}_{i1})$. From the first term of \eqref{49}:
\begin{equation}\label{15}
     \begin{split}
\ell(D_{i1}^{(m)},\pmb{\gamma}^{(m)},\pmb{y}_{i1})&=\sum_{i=1}^{n}
\left\{ D_{i1}^{(m)}\pmb{z}_{i1}'\pmb{\gamma}^{(m)}-\ln[1+\exp{(\pmb{z }_{i1}'\pmb{\gamma}^{(m)})}]\right\} w_{i1(k)}^{(m)} \\
\end{split}
\end{equation}
where $w_{i1(k)}^{(m)}$ is the corresponding weight for the $i\text{th}$ observation at the $m\text{th}$ iteration of the algorithm at time 1 {in the $k\text{th}$ possible value of the response variable}. The proposal of \cite{costa2003modelos} is used, which uses the estimated values in step $E$ of the indicator variable, as the response variable in maximizing the likelihood for the estimation of the parameter $\pmb{\gamma}$. To do this, first calculate the partial derivative of \eqref{15} with respect to $\pmb{\gamma}^{(m)}$, which is equivalent to:
\begin{equation*}
\begin{split}
\dfrac{\partial \ell(D_{i1}^{(m)},\pmb{\gamma}^{(m)},\pmb{y}_{i1})}{\partial \pmb{\gamma}'}&=\sum_{i=1}^{n}\left\{ D_{i1}^{(m)}\pmb{z}_{i1}'-\dfrac{\exp{(\pmb{z}_{i1}' \pmb{\gamma}^{(m)})}\pmb{z}_{i1}'}{1+\exp{(\pmb{z}_{i1}'\pmb{\gamma}^{(m)})}}
\right\}w_{i1(k)}^{(m)} \\
&=\sum_{i=1}^{n} \pmb{z}_{i1}'\left[ D_{i1}^{(m)}-\dfrac{\exp{(\pmb{z}_ {i1}'\pmb{\gamma}^{(m)})}}{1+\exp{(\pmb{z}_{i1}'\pmb{\gamma}^{(m)})} }\right] w_{i1(k)}^{(m)} \\
&=\sum_{i=1}^{n} \pmb{z}_{i1}'\left[ D_{i1}^{(m)}-\pi_{i1}^{(m)}\right ]w_{i1(k)}^{(m)}
\end{split}
\end{equation*}
and the second derivative of \eqref{15} with respect to $\pmb{\gamma}^{(m)}$ is:
\begin{align}\label{53}
\dfrac{ \partial^{2} \ell(D_{i1}^{(m)},\pmb{\gamma}^{(m)},\pmb{y}_{i1}) }{\partial \pmb{ \gamma} \partial\pmb{ \gamma}'}&=-
\sum_{i=1}^{n}\pmb{z}_{i1}'\left[\dfrac{\exp{(\pmb{z}_{i1}' \pmb{\gamma}^{ (m)})} }{[1+\exp{(\pmb{z}_{i1}'\pmb{\gamma}^{(m)})}]^{2}} \right]w_{ i1(k)}^{(m)} \pmb{z}_{i} \nonumber \\
&=-\sum_{i=1}^{n} \pmb{z}_{i1}'\left[ \pi_{i1}^{(m)}(1-\pi_{i1}^{(m )}) \right]w_{i1(k)}^{(m)} \pmb{z}_{i}
\end{align}
Then, the equation to solve is:
  \begin{align} \label{16}
\sum_{i=1}^{n}\left\{ D_{i1}^{(m)}\pmb{z}_{i1}'-\dfrac{\exp{(\pmb{z}_{ i1}'\pmb{\gamma}^{(m)})}\pmb{z}_{i1}'}{1+\exp{(\pmb{z}_{i1}'\pmb{\gamma }^{(m)})}}
\right\}w_{i1(k)}^{(m)} &=0
\end{align}
This deduction of the weights from the likelihood function, for the estimation of the missing data, coincides with the proposal of \cite{hall2010robust}, who assigned weights to perform a more robust zero-inflated regression.
To estimate the parameters $\pmb{\beta}$ and $\pmb{\gamma}$, the Fisher-Scoring method is used, from which it follows that, in the case of the estimate of $\pmb{\gamma}$ in the equation \eqref{16}:
\begin{equation}\label{17}
\pmb{ \gamma}^{(m+1)}=\pmb{\gamma}^{(m)}+ [\pmb{\Im}_{z}^{(m)}]^{-1}\pmb{U}_{z}^{(m)}
\end{equation}
with
\begin{align*}
    \pmb{\Im}_{z}^{(m)}&=\sum_{i=1}^{n} \pmb{z}_{i1}'\pi_{i1}^{(m)}(1-\pi_{i1}^{(m)})w_{i1(k)}^{(m)} \pmb{z}_{i1}=\pmb{Z}_{1}'\pmb{M}_{1z}^{(m)}\pmb{ W}_{1}^{(m)}\pmb{Z}_{1}
\end{align*}
where $\pmb{W}_{1}^{(m)}=diag(w_{i1(k)}^{(m)})$ and $\pmb{M}_{1z}^{(m )}=diag(\pi_{i1}^{(m)}(1-\pi_{i1}^{(m)}))$ are matrices of size $n \times n$, and $\pmb{Z}_1'$ is the matrix of covariates associated with the zero-inflated model at time 1 of size $n \times p$. In addition,
\begin{align}
     \pmb{U}_{z}^{(m)}&=\sum_{i=1}^{n} \pmb{z}_{i1}'\left[D_{i1}^{(m) }-\pi_{i1}^{(m)} \right] w_{i1(k)}^{(m)}\nonumber \\
     &=\sum_{i=1}^{n}\left\{ \pmb{z}_{i1}'\left[D_{i1}^{(m)}-\pi_{i1}^{(m )} \right] w_{i1(k)}^{(m)} \dfrac{\pi_{i1}^{(m)}(1-\pi_{i1}^{(m)})}{\pi_{i1}^{(m)}(1-\pi_{i1}^{(m)}) } \right\}\label{19}
\end{align}
{
when replaced in \eqref{19}, $ \pmb{U}_{z}^{(m)}$ in matrix form is expressed as:
\begin{equation}\label{21}
  \pmb{U}_{z}^{(m)}= \pmb{Z}_{1}' \pmb{M}_{1z}^{(m)} \pmb{W}_{1} ^{(m)} \pmb{v}_{z1}^{(m)}
\end{equation}
where $ \pmb{v}_{z1}^{(m)}$ of size $n \times 1$ is defined as:
\begin{align}
  \pmb{v}_{z1}^{(m)}&=\Bigg( \dfrac{D_{11}^{(m)}-\pi_{11}^{(m)}  }{\pi_{11}^{(m)}(1-\pi_{11}^{(m)}) }  \ldots, \dfrac{D_{i1}^{(m)}-\pi_{i1}^{(m)} }{\pi_{i1}^{(m)}(1-\pi_{i1}^{(m)}) } ,
  \ldots,   \dfrac{D_{n1}^{(m)}-\pi_{n1}^{(m)}}{\pi_{n1}^{(m)}(1-\pi_{n1}^{(m)}) } \Bigg)'\label{20}
\end{align}
}
and replacing \eqref{21} in \eqref{17}, it is obtain that:
\begin{equation*}
\pmb{\gamma}^{(m+1)}=\left[ \pmb{Z}_{1}'\pmb{M}_{1z}^{(m)}\pmb{W}_{1}^{(m)}\pmb{Z}_{1} \right]^{-1}  \pmb{Z}_{1}'\pmb{M}_{1z}^{(m)}\pmb{W}_{1}^{(m)}\left( \pmb{Z}_{1}\pmb{\gamma}^{(m)}+ \pmb{v} _{z1}^{(m)} \right) 
\end{equation*}
Similarly, $\pmb{\beta}^{(m)}$ is estimated by maximizing $\ell(D_{i1},\pmb{\beta}^{(m)},y_{i1} )$ as follows: from the first term of \eqref{52}, it is obtained that:
\begin{align*}
\ell(D_{i1}^{(m)},\pmb{\beta}^{(m)},y_{i1})&=\sum_{i=1}^{n} (1-D_{ i1}^{(m)})\left[ y_{i1}\pmb{x}_{i1}'\pmb{\beta}^{(m)}-\exp{(\pmb{x}_{ i1}'\pmb{\beta}^{(m)})} \right] w_{i1(k)}^{(m)}
\end{align*}

The partial derivative with respect to $\pmb{\beta}$ is:
\begin{align}\label{54}
\dfrac{\partial \ell(D_{i1}^{(m)},\pmb{\beta}^{(m)},y_{i1})}{\partial \pmb{\beta}'}& =
\sum_{i=1}^{n} (1-D_{i1}^{(m)})\left[ y_{i1}\pmb{x}_{i1}'-\exp{(\pmb{ x}_{i1}'\pmb{\beta}^{(m)})}\pmb{x}_{i1}' \right]w_{i1(k)}^{(m)} \nonumber \\
&=\sum_{i=1}^{n}\pmb{x}_{i1}' (1-D_{i1}^{(m)}) \left[ y_{i1}-\lambda_{i1} ^{(m)} \right] w_{i1(k)}^{(m)}
\end{align}
The second partial derivative with respect to $\pmb{\beta}$ is:
\begin{align*}
\dfrac{\partial^{2} \ell(D_{i1}^{(m)},\pmb{\beta}^{(m)},y_{i1})}{\partial \pmb{\beta } \partial \pmb{\beta}'}&=
\sum_{i=1}^{n} - \pmb{x}_{i1}' (1-D_{i1}^{(m)}) \left[\exp{(\pmb{x}_{ i1}'\pmb{\beta}^{(m)})} \right]
w_{i1(k)}^{(m)} \pmb{x}_{i1}
\end{align*}
Analogously, it is obtained that:
\begin{align*}
\pmb{\Im}_{x}^{(m)}&=\sum_{i=1}^{n} \pmb{x}_{i1}'(1-D_{i1}^{(m)}) \lambda_{i1}^{(m)} w_{ i1(k)}^{(m)} \pmb{x}_{i1} =\pmb{X_{1}}'\pmb{W}_{1}^{(m)}\pmb{M}_{1x}^{(m)}\pmb{X}_{1}
\end{align*}
where $\pmb{M}_{1x}^{(m)}=diag(\lambda_{i1})$ is a matrix of size $n \times n$ and $\pmb{X}_{1}$ is the matrix of covariates associated with the Poisson model at time 1 of size $n \times p$. In addition
\begin{align*}
\pmb{U}_{x}^{(m)}&=\sum_{i=1}^{n} w_{i1(k)}^{(m)} (1-D_{i1}^{ (m)})\pmb{x}_{i1}'\left[ y_{i1}-\lambda_{i1}^{(m)} \right]=\pmb{X}_{1}'\pmb{W}_{1}^{(m)}\pmb{v}_{x1}^{(m)}
\end{align*}
{
where
{\small
\begin{equation*}
  \pmb{v}_{x1}^{(m)}= \left( \dfrac{(1-D_{11}^{(m)}) \left[ y_{11}-\lambda_{11}^{(m)} \right]}{\lambda_{11}^{(m)}}, \ldots,  \dfrac{(1-D_{i1}^{(m)}) \left[ y_{i1}-\lambda_{i1}^{(m)} \right]}{\lambda_{i1}^{(m)}} ,\ldots,\dfrac{(1-D_{n1}^{(m)}) \left[ y_{n1}-\lambda_{n1}^{(m)} \right]}{\lambda_{n1}^{(m)}} \right)'
\end{equation*}
}
}
Similarly to the estimate of $\pmb{\gamma}^{(m+1)}$, the estimate of $\pmb{\beta}$ is given by:
\begin{align*}
\pmb{\beta}^{(m+1)}&=[\pmb{X}_{1}'\pmb{W}_{1}^{(m)}\pmb{M}_{1x}^{(m)}\pmb{X}_{1}]^{-1} \pmb{X}_{1}'\pmb{W}_{1}^{(m)}\pmb{M}_{1x}^{(m)}\left( \pmb{X}_{1}\pmb{\beta}^{(m)}+ \pmb{v}_{x1}^{(m)}\right)
\end{align*}
Once $\pmb{\beta}$ and $\pmb{\gamma}$ are estimated, then in Step 2, the missing data at time 1 are estimated.
\subsubsection{Step 2}
Estimation and imputation of the missing data at time 1 are developed in this subsection. The EM algorithm is performed for the data at time $t=1$ taking into account the parameters obtained in step 1 as the initial estimator.
\paragraph{Step E.} Step E imputes the values for the missing data using the model proposed by \cite{bartlett1937some}:
\begin{align}\label{22}
    E[g(\pmb{y}_{1})\vert \pmb{X}_{1},\pmb{Z}_{1}]=
\begin{pmatrix}
\pmb{X}_{1} \\
\pmb{0}
\end{pmatrix}
  \pmb{\beta}
+
\begin{pmatrix}
\pmb{0} \\
\pmb{Z}_{1}
\end{pmatrix}
  \pmb{\gamma}
+
\begin{pmatrix}
\pmb{A}_{1} \\
\pmb{0}
\end{pmatrix}
  \pmb{\alpha}
+
\begin{pmatrix}
\pmb{0} \\
\pmb{B}_{1}
\end{pmatrix}
  \pmb{\tau}
\end{align}
where $\pmb{A}_{1}$ and $ \pmb{B}_{1}$ are matrices of sizes $n \times (n-n_{1})$, $n_{1}$ indicates the total data observed at time 1, and $\pmb{A}_{1}$ and $ \pmb{B}_{1}$ correspond to the covariates of the Poisson zero-inflated models, respectively. $\pmb{\alpha}$ and $\pmb{\tau}$ correspond to the vectors of the $(n-n_{1})$ regression coefficients for the missing covariate value, and $\pmb{\beta} $ and $\pmb{\gamma}$ are the coefficients estimated in Step 1. The matrices $\pmb{X}_{1}$ and $ \pmb{Z}_{1}$ are the associated covariate matrices at time $t=1$ of the Poisson model and inflated zero, respectively.
The matrices $\pmb{A}_{1}$ and $ \pmb{B}_{1}$ are partitioned between observed and expected values, and it is obtained as:

{\footnotesize
\begin{align} \label{23}
     \begin{pmatrix}
\ln(\lambda_{1(nzo)})\\
\ln(\lambda_{1(nzm)})\\
\ln\left( \frac{\pi_{1(zo)}}{1-\pi_{1(zo)}} \right)   \\
\ln\left( \frac{\pi_{1(zm)}}{1-\pi_{1(zm)}}  \right)
\end{pmatrix}
=\begin{pmatrix}
\pmb{X}_{1(obs)} \\
\pmb{X}_{1(miss)} \\
\pmb{0} \\
\pmb{0} \\
\end{pmatrix}
\pmb{\beta}
+
\begin{pmatrix}
\pmb{0} \\
\pmb{0} \\
\pmb{Z}_{1(o)} \\
\pmb{Z}_{1(m)}
\end{pmatrix}
\pmb{\gamma}
+
\begin{pmatrix} \pmb{0}_{1(obs)} \\
- \pmb{I}_{1(miss)} \\
\pmb{0}\\
\pmb{0}
\end{pmatrix}
\pmb{\alpha}
+
\begin{pmatrix}
\pmb{0} \\
\pmb{0} \\
\pmb{0}_{1(obs)} \\
-\pmb{I}_{1(miss)}
\end{pmatrix}
\pmb{\tau}
\end{align}
}
where $\pmb{X}_{1(obs)}$ and $\pmb{Z}_{1(obs)}$ correspond to matrices of observed covariate values, $\pmb{X}_{1(miss)}$ and $\pmb{Z}_{1(miss)}$ correspond to covariate matrices of missing values at time 1, both from the Poisson model and the Zero model inflated, respectively, $\pmb{0}_{1(obs)}$ are matrices related to the observed data, $\pmb{I}_{1(miss)}$ are identity matrices related to the missing data at time 1, $\pmb{\lambda}_{1(nzo)}$ is the vector related to the non-zero observed values at time 1, $\pmb{\lambda}_{1(nzm)}$ is the vector related to the non-zero missing values at time 1, $\pmb{\lambda}_{1(zo)}$ is the related vector of the observed values that are zero at time 1, and $\pmb{\lambda}_{1(zm)}$ is the vector of missing values that are zero at time 1.

Based on the above, for time $t=1$, the expected $\log$-$\text{likelihood}$ given the observed data is written as:
\begin{align}\label{24}
Q_{i1}[\pmb{\theta}\vert \pmb{\theta}^{(m)}]&=E[L(\pmb{\theta} ;g(y_{i1}))\vert \pmb{x}_{i1},\pmb{z}_{i1},\pmb{y}_{i1(obs)}, \pmb{\theta}=\pmb{\theta}^{(m)}]
\end{align}
where $\pmb{\theta}=(\pmb{\beta}, \pmb{\gamma},\pmb{ \alpha},\pmb{ \tau})$. From \eqref{24}, it is obtained that:
\begin{equation} \label{25}
    \begin{split}
E[Y_{1(obs)}]&=\pmb{X}_{1(obs)} \pmb{\beta}+  \pmb{Z}_{1(obs)}    \pmb{\gamma}  \\
E[Y_{1(miss)}]&= \pmb{X}_{1(miss)}\pmb{\beta}-\pmb{\alpha}+\pmb{Z}_{1(miss)}\pmb{\gamma}-\pmb{\tau}
    \end{split}
\end{equation}

Assuming an initial vector for $Y_{1(miss)}$, $\hat{Y}_{1(miss)}=0$, according to \cite{bartlett1937some} and considering that the model that concerns the Poisson part is independent of the model of zeros, the sum of the squares of the errors to be minimized over $\pmb{\theta}^{ (m)}$ is:
\begin{align} \label{55}
SCE(\pmb{\theta}^{ (m)})&=\sum_{i=1}^{n_{nzo}}({y}_{i1nzo}-\pmb{x}_{i1 (obs)}'\pmb{\beta})^{2}
+\sum_{i=n_{nzo}+1}^{n_{nzm}+n_{nzo}}({\widehat{y}}_{i1nzm}-\pmb{x}_{i1(miss)}'\pmb {\beta} +\pmb{\alpha})^{2} \nonumber \\
&+
\sum_{i=1}^{n_{zo}}({y}_{i1zo}-\pmb{z}_{i1(obs)}'\pmb{\gamma})^2+
\sum_{i=n_{zo}+1}^{n_{zo}+n_{zm}}({\widehat{y}}_{i1zm}-\pmb{z}_{i1(miss)}'\pmb{ \gamma} +\pmb{\tau})^{2}
\end{align}
where $n_{zo}$ refers to the number of observed values that are zero, $n_{zm}$ to the number of missing values that are zero, $n_{nzo}$ to the number of observed values that are not zero and $m_{nzm}$ to the number of missing non-zero values. Considering the initial assumption about $\pmb{\widehat{y}}_{i1(miss)}=0$, the least squares estimator for $\pmb{\alpha}$ and $\pmb{\tau }$, gets from \eqref{55}, first derives and gets:
\begin{align*}
\dfrac{\partial SCE(\pmb{\theta}^{(m)})}{\partial \pmb{\alpha}}= 2\sum_{i=n_{nzo}+1}^{n_{nzo}+n_{nzm}}( \widehat{y}_{i1nzm}-\pmb{x}_{i1}'\pmb{\beta}+\pmb{\alpha})&=0 \\
\dfrac{\partial SCE(\pmb{\theta}^{(m)})}{\partial \pmb{\tau}}= 2\sum_{i=n_{zo}+1}^{n_{zo}+n_{zm}}( \widehat{y}_{i1zm}-\pmb{z}_{i1}'\pmb{\gamma}+\pmb{\tau})&=0 \end{align*}
Solving the equations and writing in matrix form, it is obtained that:
\begin{align*}
\widehat{\pmb{\alpha}}&=\pmb{X}_{1(miss)}\widehat{\pmb{\beta}}\\
     \widehat{\pmb{\tau}}&=\pmb{Z}_{1(miss)}\widehat{\pmb{\gamma}}
\end{align*}
Therefore, the least squares estimator for the missing information is obtained in an iterated form using the following equations:
\begin{equation*}
\begin{cases}
\widehat{\alpha}^{(m)}_{i}=\pmb{x}_{i1(miss)}'\widehat{\pmb{\beta}}^{(m)}=\ln[\widehat{\lambda}_{i1(miss)}^{(m)}]  &\text{if} \ y_{i1} > 0, \ i=n_{nzo}+1,\ldots, n_{nzo}+n_{nzm}   \\
\widehat{\tau}^{(m)}_{i}=\pmb{z}_{i1(miss)}'\widehat{\pmb{\gamma}}^{(m)}=\ln\left[\dfrac{\widehat{\pi}_{i1(miss)}^{(m)}}{1-\widehat{\pi}_{i1(miss)}^{(m)}}\right] &\text{if} \ y_{i1}= 0, \  i=n_{zo}+1,\ldots, n_{zm}+n_{zo}
\end{cases}
\end{equation*}
Therefore,
\begin{align*}
\widehat{\lambda}_{i1(miss)}^{(m)}&= \exp{(\pmb{x}_{i1(miss)}'\pmb{\beta}^{(m)})}&\text{if} \ y_{i1}  > 0 \\
\widehat{\pi}_{i1(miss)}^{(m)} &=\dfrac{\exp{(\pmb{z}_{i1(miss)}'\pmb{\gamma}^{(m)})}}{1+\exp{(\pmb{z}_{i1(miss)}'\pmb{\gamma}^{(m)})}}  &\text{if} \  y_{i1} = 0
\end{align*}
where $\hat{\lambda}_{i1}^{(m)}$ and $\hat{\pi}_{i1}^{(m)}$ refer to the value of the mean corresponding to the individual $i $ at time 1 if the answer is different from zero or equal to zero, respectively. Given the $m$th iteration, the selection criteria is:
\begin{center}
$\widehat{y}_{i1(miss)} = \begin{cases}
0&\text{if } \ \widehat{\pi}_{i1(miss)}^{(m)} >p_{o} \\
   \widehat{\lambda}_{i1(miss)}^{(m)}&\text{if } \ \widehat{\pi}_{i1(miss)}^{(m)}\leq p_{o }\\
\end{cases} $
\end{center}
The value $p_{0}$ is a reference value to impute the data as zero or different from zero. In inflated zero variables, generally according to \cite{costa2003modelos}, the percentage of zeros is at least $40\%$ of the observations. Therefore, 0.4 may be a minimum value if there is no in-depth knowledge of the variable being worked on.
\paragraph{Step M.} Once the missing observations are imputed, the maximization will be carried out starting from the complete data set and using the Fisher-Scoring method, with the initial estimator given in Step 1. Steps E and M are carried out until the convergence. The indicator variable that allows estimating the excess of zeros and the Poisson model as independent is a vector of size $n$, which takes values according to:
\begin{align*}
E\left[D_{i1}^{(m)}\vert y_{i1},\pmb{\gamma}^{(m)},\pmb{\beta}^{(m)} \right]& = \begin{cases}
\dfrac{1}{1+ \exp{(-\pmb{z}_{i1}'\pmb{\gamma}^{(m)}-\exp{(\pmb{x}_{i1}' \pmb{\beta}^{(m)})})} } &\text{if } \ y_{i1}=0 \\
0&\text{if } \ y_{i1}> 0\end{cases}
\end{align*}
The expressions for the estimation of the parameters are:
\begin{align*}
\pmb{\gamma}^{(m+1)}&=\left[ \pmb{Z}_{1}'\pmb{M}_{1z}^{(m)}\pmb{Z}_ {1} \right]^{-1} \pmb{Z}_{1}'\pmb{M}_{1z}^{(m)}\left( \pmb{Z}_{1}\pmb {\gamma}^{(m)}+ \pmb{v}_{z1}^{(m)} \right) \\
\pmb{\beta}^{(m+1)}&=[\pmb{X}_{1}'\pmb{M}_{1x}^{(m)}\pmb{X}_{1 }]^{-1} \pmb{X}_{1}'\pmb{M}_{1x}^{(m)}\left( \pmb{X}_{1}\pmb{\beta} ^{(m)}+ \pmb{v}_{x1}^{(m)}\right)
\end{align*}
with $\pmb{Z}_{1}$ and $\pmb{X}_{1}$ of sizes $n\times p$. $\pmb{M}_{1z}^{(m)}=diag(\pi_{i1}^{(m)}(1-\pi_{i1}^{(m)}))$ and $\pmb{M}_{1x}^{(m)}=diag(\lambda_{i1}^{(m)})$ of sizes $n \times n$.
Once finished the complete imputation of the data for time $t=1$ and the estimation of $\pmb{\beta}$ and $\pmb{\gamma}$, the imputation of the data for the time $t=2$, will be done as explained in the next section.
\subsection{Time 2}
In the estimation of the model parameters for the second time, the response variable of the time $t=1$, $y_{i1}$ is used as a covariate associated with the model using the $EM$ algorithm with weights. During the imputation of the missing values for the time $t=1$, all the entries of the response variable may be zero, in which case imputing the data for the time $t=1$ does not improve the explanation of the variability of the response variable. Therefore, if this case happens, this variable should not be considered and the link functions for the parameters $\lambda_{i2}$ and $\pi_{i2}$ should be defined as:
\begin{align}
     \ln[\lambda_{i2}]=\pmb{x}_{i2}'\pmb{\beta} &\qquad \text{if} \ y_{i2} > 0 \nonumber \\
\ln\left[\dfrac{\pi_{i2}}{1-\pi_{i2}}\right]=\pmb{z}_{i2}'\pmb{\gamma} &\qquad \text{if} \ y_{i2} = 0   \label{26}  
\end{align}

The estimation of the model parameters and the imputation of the data is carried out in a similar way as proposed for the time $t=1$. But if the inputs of $y_{i1}$ are not all zero, which is the most probable situation, the estimation of the vector of maximum plausible parameters in time $t=2$ must be carried out in the same way as in Step 1 of time $ t=1$, with the use of the $EM$ algorithm taking into account $y_{i1}$ as a covariate associated in the model, that is, the link functions for the parameters are as in \eqref{26}. Those can be written in matrix form as:
\begin{align*}
    E[ g(\pmb{y}_{2})]= \begin{pmatrix}
\pmb{X}_{2} &\pmb{Z}_{2} \end{pmatrix}
\begin{pmatrix}
\pmb{\beta} \\
\pmb{\gamma} \\
\end{pmatrix}
+ \pmb{e}
\end{align*}
where $\pmb{X}_{2}$ and $\pmb{Z}_{2}$ are the matrices of size $n\times (p+1)$ of covariates related to $\pmb{y}_ {2}$, which is the response vector at time $t=2$. There are $p$ columns that contain the independent variables that explain the behavior of the dependent variable and the last column is the response vector at time $t= $1 including the imputed value. The complete deduction for the time $t=2$ is shown in the appendix \ref{ApenAA}.
Now, the proposed algorithm for time $t$ will be shown.
\subsection{Time $t$}
The estimation of the vector of maximum plausible parameters in time $t$ is carried out using the EM algorithm with weights. Let $\pmb{Y}_{1},\pmb{Y}_{i2},\ldots,\pmb{Y}_{(t-1)}$ be the response vectors from time $1$ to time $t-1$ obtained by the previous steps. To avoid multicollinearity problems due to the longitudinal measurement {and to take into account the correlation that may exist between the same observations of an experimental unit, i.e. the correlation that exists between $\pmb{Y}_{t'}$ and $\pmb{Y}_{t}$}, the PCA method is used to generate a new set of representative orthogonal covariates from the values $\pmb{Y}_{1},\pmb{Y}_{i2},\ldots,\pmb{Y}_{(t-1)}$. The covariate matrix is then given by:
\begin{equation*}
  \pmb{C}_{t-1}=[\pmb{Y}_{1},\pmb{Y}_{2},\ldots,\pmb{Y}_{(t-1)}]\pmb{A}'
\end{equation*}
where $\pmb{A}=[\pmb{a}_{1},\pmb{a}_{2},\ldots,\pmb{a}_{t-1}]$ is the matrix of eigenvectors corresponding to the eigenvalues of the PCA performed on $\pmb{Y}_{1},\pmb{Y}_{i2},\ldots,\pmb{Y}_{(t-1)}$. The full deduction for time $t$ is done similarly to the algorithm for time $t=2$, and it is shown in the appendix \ref{ApenAB}. 
In the next section, the effectiveness of the method will be tested to reconstruct missing information and improve parameter estimation through simulation.
\section{Simulation}
\begin{figure}
    \centering
    \includegraphics[width=14cm]{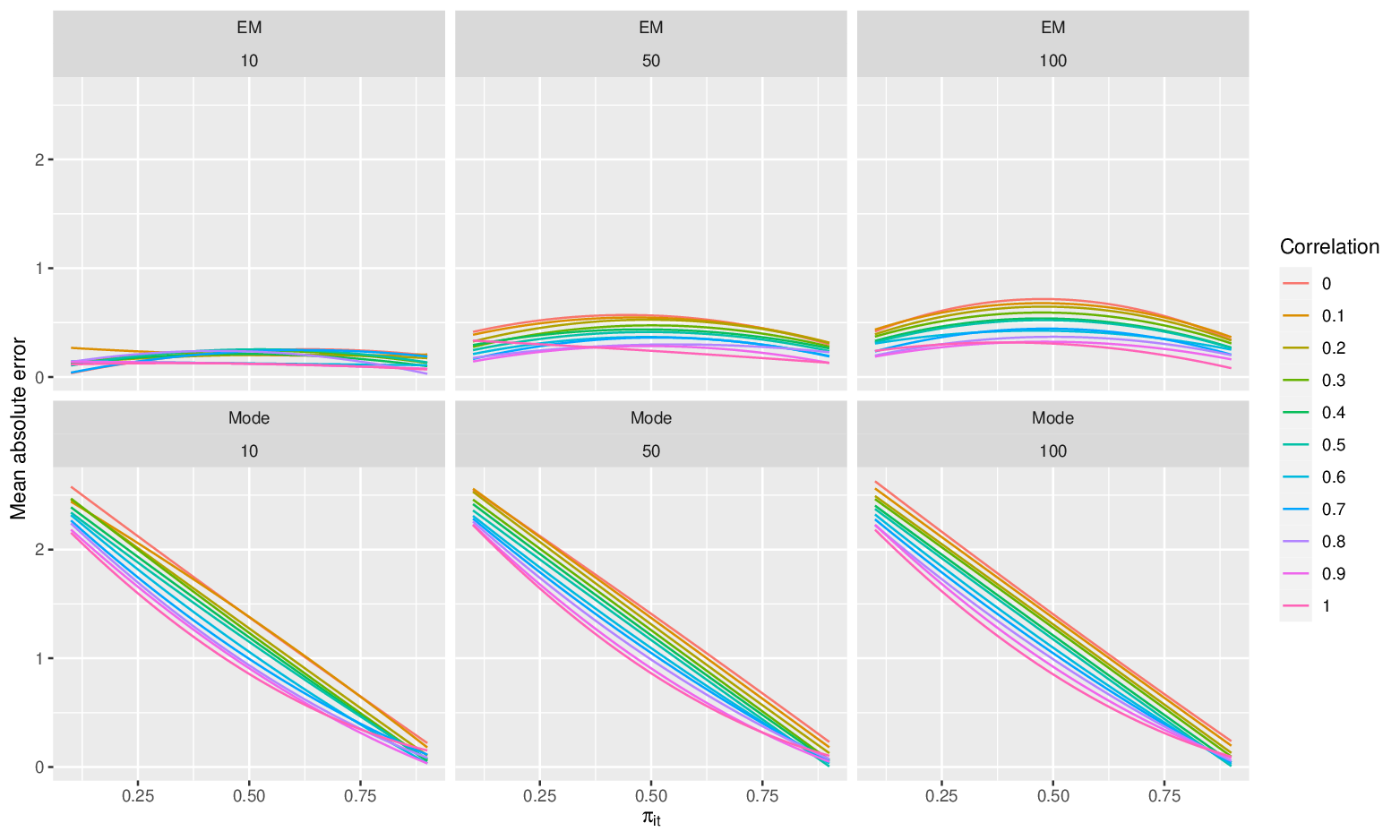}
    \caption{Mean absolute error between the real value and the value filled by two models: EM) the methodology proposed in this work and Mode) the mode of the individual's values over time. $\pi_{it}$ is the probability of zero associated with the ZIP model and the correlation is the value of $\alpha$ in the correlation matrix.}
    \label{fig1}
\end{figure}

{
For the simulation exercise, three different treatments are assumed: A, B, and C. Each one is applied to $n$ experimental units during $T$ different times. The random variable $Y_{ij}$ is assumed to have a zero-inflated Poisson distribution with the following models:
\begin{align*}
\ln(\lambda _{it})&=\beta_0+\beta_1x_{i1}+\beta_2x_{i2} + \beta_3 t,\qquad \ln \left(\dfrac{\pi _{it}}{1-\pi _{it}} \right)=\gamma_0\\
i&=1, \ldots, 3n \qquad t=1, \ldots, T \qquad Cor(\pmb{Y}_i)=R(\alpha)
\end{align*}
where $x_{i1}$ is a variable that has a value of 1 if treatment B is applied to the experimental unit $i$, and $x_{i2}$ is a variable that has a value of 1 if treatment C is applied to the experimental unit $i$. The values defined for the simulation parameters were: $\beta_0=1$, $\beta_1=-0.5$, $\beta_2=0.5$ and $\pi_{it}\in\{0.1, \ldots, 0.8\}$, $T\in\{5,10\}$, $n\in\{10, 50, 100\}$ . Furthermore, the correlation between the observations of an experimental unit is modeled using an autoregressive correlation of order 1 or an exchangeable one and defined as $R(\alpha)$, with a value of $\alpha\in\{0, 0.1, \ldots, 0.9\}$. 
\\
Each simulation was run 1000 times and the following procedure was carried out: i) complete model: A ZIP model was run with the complete data, ii) information was eliminated in percentages from 0.1 to 0.8, iii) missing model: a ZIP model was fitted but only with complete data, iv) mode model: the missing data were filled with mode of each individual and a ZIP model was fitted and v) EM model; The data was filled with the methodology proposed in the previous sections and the ZIP model was fitted. With the Mode and EM models, the average absolute difference between the real value and the one filled with the model was calculated. The results are shown in Figure \ref{fig1}.
}
\\
{
In Figure \ref{fig1}, it is observed that the methodology proposed in this paper has a better performance than the individual's Mode regardless of the proportion of zeros or the correlation between the individuals. Furthermore, it is highlighted that as the correlation increases, the distance from the real value is even smaller regardless of the proportion of zeros in the random variable. Note in Figure \ref{fig1} that if the proportion of zeros is smaller, the model with Mode has a worse performance than if the proportion of zeros is larger; a similar behavior regardless of the number of individuals per sequence. This is explained because increasing the probability of zero makes it easier to fill the data with values close to 0. Furthermore, when differentiating by loss percentages, the advantage of the methodology proposed in this paper is even more noticeable when the loss percentage is high. This is observed in the figures presented in the supplementary file \ref{SF1}.}
\begin{figure}
    \centering
    \includegraphics[width=14cm]{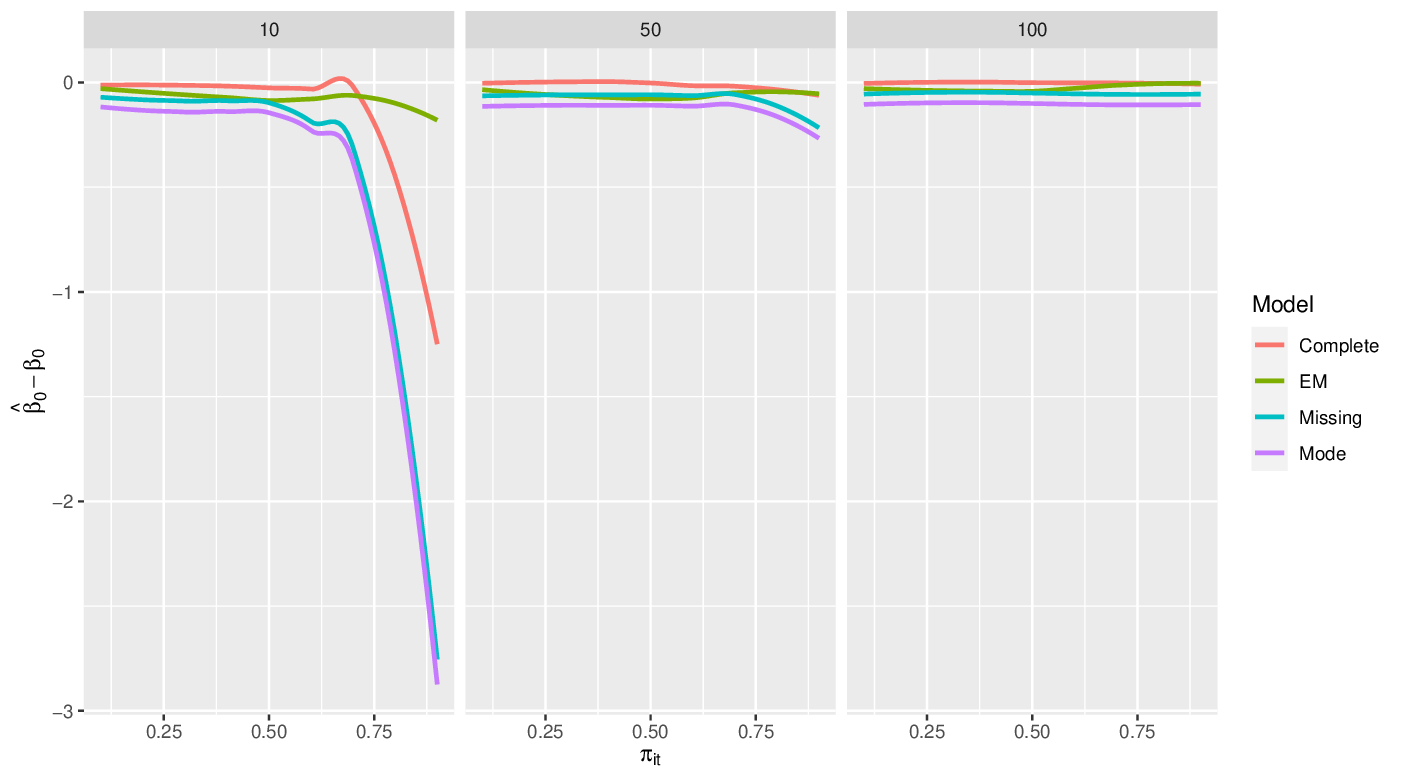}
    \caption{Average bias $\hat{\beta}_0-\beta_0$ for each of the models according to the number of individuals receiving treatment. $\pi_{it}$ is the probability of zero associated with the ZIP model and the average value over all percentages of missing data is presented.}
    \label{fig2}
\end{figure}
\\
{It is observed in Figure \ref{fig2} that the estimation bias of $\beta_0$ decreases as the number of individuals per sequence increases, and when $n=10$, the model with the data filled by the methodology proposed in This work has less bias compared to models that fill in with the mode, or that eliminate missing information. 
\begin{figure}
    \centering
    \includegraphics[width=14cm]{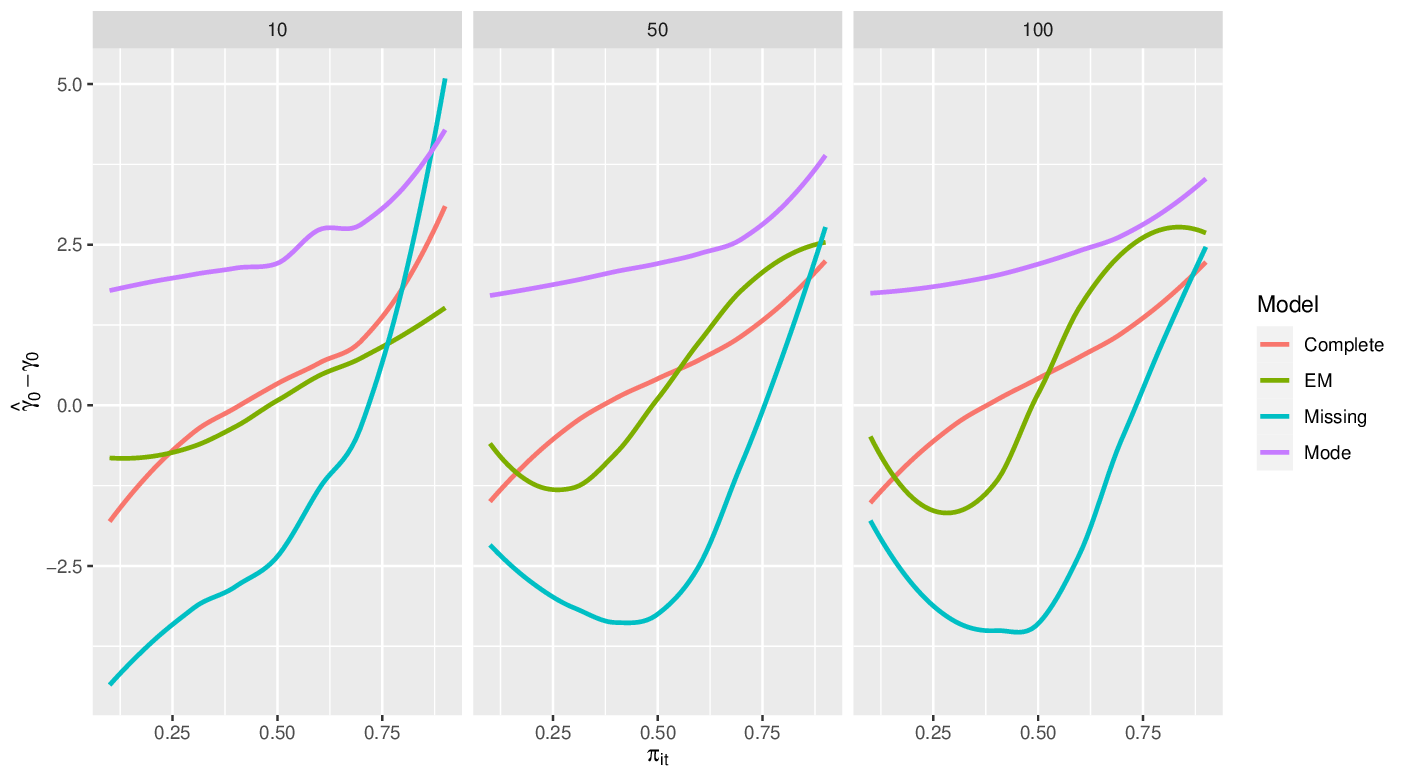}
    \caption{Average bias $\hat{\gamma}_0-\gamma_0$ for each of the models according to the number of individuals receiving treatment. $\pi_{it}$ is the probability of zero associated with the ZIP model and the average value over all percentages of missing data is presented.}
    \label{fig3}
\end{figure}
\\
This behavior is also reflected in Figure \ref{fig3} where it is also shown that the estimation of the proportion of zeros has a very large bias in models that do not treat the missing information adequately. This demonstrates the good performance of the proposed methodology to fill in the missing data and also allows us to obtain estimates with a lower bias than ignoring the missing data or filling them with the mode. When the figures are analyzed by percentage of loss in the supplementary file \ref{SF1}, the behavior is similar to that presented in figures \ref{fig2} and \ref{fig3}.
With all of the above, it is highlighted that the ZIP model is sensitive to the loss of information and without treating the missing data there may be a bias in the inflated zero part of the model. Furthermore, if they are replaced by the mode, the estimate of $\gamma_0$ is significantly underestimates the parameter, which is explained by the large number of 0s that could appear in the filled data.
Furthermore, when working with higher correlation values between individuals, the estimation with the EM model improves significantly, as seen in the figures in the supplementary material \ref{SF1}.}

Now, the algorithm is used in a real data set to check its effectiveness in different scenarios of information loss.
\section{Application}
This section shows the methods for the estimation and imputation of missing information. The application presented is taken from a study of maize genetic improvement, this data was worked by \cite{costa2003modelos}. 
The experiment began on March 11, 2001, and had the objective of evaluating the efficiency of the genetically modified maize MON810 with conventional maize (hybrid DKB909) in the control of \text{Spodoptera} \text{frugiperda}. The experiment was completely randomized, with 3 treatments and 8 replications in plots of $1250m^{2}$, being evaluated for 9 weeks. The treatments were:
\begin{enumerate}
     \item Genetically modified maize MON810, treatment 1.
     \item Conventional corn with application of insecticides, treatment 2.
     \item Conventional corn without application of insecticides, treatment 3.
\end{enumerate}
In this application, it was established that whenever $20\%$ to $30\%$ of the plants had symptoms of attacks, the insecticide would be applied. Table \ref{origin1} shows the data from the maize study, with measurements over nine weeks of the number of large caterpillars in each of the eight plots that were part of the study.
\begin{table}[ht]
\centering
\begin{tabular}{r|rrrrrrrrrr}
   \hline
  & We1 & We2 & We3 & We4 & We5 & We6 & We7 & We8 & We9 & Treat \\
   \hline
1 & 0 & 0 & 0 & 0 & 0 & 0 & 0 & 0 & 0 & 1 \\
   2 & 0 & 0 & 0 & 0 & 0 & 0 & 0 & 0 & 0 & 1 \\
   3 & 0 & 0 & 0 & 0 & 0 & 0 & 0 & 0 & 0 & 1 \\
   4 & 0 & 0 & 0 & 0 & 0 & 0 & 0 & 0 & 1 & 1 \\
   5 & 0 & 0 & 0 & 0 & 0 & 0 & 0 & 0 & 1 & 1 \\
   6 & 0 & 0 & 0 & 0 & 0 & 0 & 0 & 0 & 1 & 1 \\
   7 & 0 & 0 & 0 & 0 & 0 & 1 & 0 & 1 & 2 & 1 \\
   8 & 0 & 0 & 0 & 0 & 0 & 1 & 0 & 1 & 3 & 1 \\
   9 & 0 & 0 & 0 & 0 & 0 & 0 & 0 & 0 & 0 & 2 \\
   10 & 0 & 0 & 0 & 0 & 0 & 0 & 0 & 0 & 0 & 2 \\
   11 & 0 & 0 & 0 & 0 & 0 & 0 & 0 & 0 & 0 & 2 \\
   12 & 0 & 0 & 0 & 0 & 0 & 0 & 0 & 1 & 0 & 2 \\
   13 & 0 & 0 & 0 & 1 & 1 & 1 & 0 & 1 & 0 & 2 \\
   14 & 0 & 0 & 0 & 0 & 1 & 2 & 1 & 1 & 0 & 2 \\
   15 & 0 & 0 & 0 & 0 & 1 & 2 & 1 & 2 & 1 & 2 \\
   16 & 0 & 0 & 0 & 0 & 2 & 4 & 1 & 2 & 3 & 2 \\
   17 & 0 & 0 & 0 & 0 & 1 & 4 & 3 & 2 & 2 & 3 \\
   18 & 0 & 0 & 0 & 0 & 1 & 5 & 4 & 2 & 3 & 3 \\
   19 & 0 & 0 & 0 & 0 & 0 & 5 & 4 & 2 & 3 & 3 \\
   20 & 0 & 0 & 0 & 0 & 0 & 5 & 5 & 2 & 4 & 3 \\
   21 & 0 & 0 & 0 & 0 & 0 & 4 & 5 & 3 & 4 & 3 \\
   22 & 0 & 0 & 0 & 0 & 0 & 8 & 6 & 3 & 6 & 3 \\
   23 & 0 & 0 & 0 & 0 & 0 & 8 & 7 & 4 & 4 & 3 \\
   24 & 0 & 0 & 0 & 0 & 0 & 9 & 7 & 4 & 4 & 3 \\
    \hline
\end{tabular}
\caption{Corn Data}\label{origin1}
\end{table}
The Poisson model and Zero excess model obtained that best fits the complete data are, respectively:
\begin{align*}
\ln(\lambda_{it})&=\beta_{0}+\beta_{1}x_{i1}+ \beta_{2}x_{i2} +\beta_{3}t \\
\ln\left(\dfrac{\pi_{it}}{1-\pi_{it}}\right)&=\gamma_{0}+\gamma_{1}x_{i1}+\gamma_{2}x_{i2}+\gamma_{3}t
\end{align*}

where $x_{i1}$ is 1 if the treatment received by the experimental plots is the second treatment and $x_{i2}$ is 1 if the treatment received by the experimental plots is the third treatment, and $t$ is a variable that measures the effect of the nine weeks of study. The estimated coefficients for both the Poisson model and the zero model with their respective $p$ values are presented in Table \ref{pc1}.

\begin{table}[ht]
\centering
\begin{tabular}{|r|rrrr|}
\hline
  \multicolumn{5}{|c|}{Poisson model} \\
\hline
            & Estimate & Std. Error &z value &$Pr(>|z|)$ \\
             \hline
$\beta_0$ & -0.525 & 0.563 & -0.931 & 0.352 \\ 
  $\beta_1$ & 0.499 & 0.373 & 1.339 & 0.181 \\ 
 $\beta_2$ & 2.326 & 0.328 & 7.081 & 0.000 \\ 
  $\beta_3$ & -0.050 & 0.059 & -0.843 & 0.399 \\ 
    \hline
  \multicolumn{5}{|c|}{Zero excess model} \\
    \hline
            & Estimate &Std. Error &z value & $Pr(>|z|)$\\
             \hline
$\gamma_0$ & 62.847 & 217.411 & 0.289 & 0.773 \\ 
  $\gamma_1$ & -19.147 & 72.491 & -0.264 & 0.792 \\ 
  $\gamma_2$ & -8.665 & 36.268 & -0.239 & 0.811 \\ 
  $\gamma_3$ & -10.619 & 36.240 & -0.293 & 0.007 \\ 
    \hline
    \end{tabular}
\caption{Parameter estimators for full data using Poisson model}\label{pc1}
\end{table}

Table \ref{pc1} shows that the third treatment effect is statistically significant for counting the number of larvae in the Poisson model. That is, the treatment influences the number of larvae found in the corn fields. The zero model is significantly explained by time (weeks). That is, the number of fields with zero larvae depends on the time in weeks. These results agree with those observed in the data.

\subsection{Estimation and imputation of missing data}
Now, to measure the effectiveness of the proposed model to complete the missing information in the different simulation scenarios with $20\%$, $30\%$, $40\%$, and $50\%$ of lost data, the parameters of the linear mixed-effect model with the complete data at the moment the missing information is imputed and confidence intervals at $95\%$ are constructed for the fixed effects.
First, random data loss is performed in percentages of $20\%$, $30\%$, $40\%$, and $50\%$, each scenario repeated 100 times. The process of estimation and imputation of the missing information proposed above is followed with zero-inflated Poisson responses, in each of the nine times. Using the algorithm programmed in \cite{RR}, the number of imputed responses is compared with the original responses, and a reason for the success of the algorithm is made. Then, a longitudinal mixed-effects model with a zero-inflated Poisson response variable is estimated and compared with the estimates from the complete data.

As an illustration, the complete result of the estimators obtained from the algorithm is shown in the appendix \ref{ApenAC} in the case that the data loss was $20\%$. Moreover, the step-by-step development of the same example with the tables, the estimated weights, and the imputed data are shown in Supplementary File 1. The two estimation steps are shown for each week: i) first step: Parameters estimated and weights, and ii) second step: imputation of loss information. Therefore, there are 18 steps in total for the estimation and imputation data.
\begin{table}[ht]
\centering
\begin{tabular}{|r|r|rrrr|rrrr|}
\hline
Percent & Coefficients & \multicolumn{4}{c|}{Poisson model} & \multicolumn{4}{c|}{Zero excess model} \\
\hline
\multirow{4}{*}{ $20\%$ }&& $\beta_0$& $\beta_1$ & $\beta_2$ & $\beta_3$  & $\gamma_0$ & $\gamma_1$ &$\gamma_2$ & $\gamma_3$ \\
  & Estimated &0.018 & 0.190 & 2.019 & -0.081 & 52.911 & -17.367 & -10.233 & -8.265\\
&Lower &-0.721 & -0.387 & 1.403 & -0.118 & 22.499 & -21.078 & -11.286 & -11.971 \\
&Upper &0.793 & 0.656 & 2.549 & -0.040 & 81.276 & -9.105 & -6.709 & -2.936 \\
\hline
& & \multicolumn{4}{c|}{Poisson model} & \multicolumn{4}{c|}{Zero excess model} \\
\hline
\multirow{4}{*}{  $30\%$ }&& $\beta_0$& $\beta_1$ & $\beta_2$ & $\beta_3$  & $\gamma_0$ & $\gamma_1$ &$\gamma_2$ & $\gamma_3$ \\
  & Estimated  & 0.091 & 0.216 & 2.021 & -0.093 & 57.566 & -17.561 & -9.330 & -9.243\\
&Lower &-0.536 & -0.387 & 1.411 & -0.143 & 22.421 & -24.763 & -10.781 & -14.474  \\
&Upper& 0.744 & 0.800 & 2.528 & -0.041 & 86.631 & -8.266 & -5.372 & -2.837\\
\hline
& & \multicolumn{4}{c|}{Poisson model} & \multicolumn{4}{c|}{Zero excess model} \\
\hline
\multirow{4}{*}{ $40\%$ }&& $\beta_0$& $\beta_1$ & $\beta_2$ & $\beta_3$  & $\gamma_0$ & $\gamma_1$ &$\gamma_2$ & $\gamma_3$ \\
  &Estimated  &0.174 & 0.243 & 1.933 & -0.091 & 52.072 & -17.061 & -10.031 & -8.088 \\
&Lower  &-0.728 & -0.263 & 1.363 & -0.148 & 20.842 & -22.701 & -11.094 & -12.834 \\
&Upper &1.061 & 1.045 & 2.624 & -0.027 & 83.342 & -8.030 & -5.542 & -2.645 \\
\hline
& & \multicolumn{4}{c|}{Poisson model} & \multicolumn{4}{c|}{Zero excess model} \\
\hline
\multirow{4}{*}{  $50\%$ }&& $\beta_0$& $\beta_1$ & $\beta_2$ & $\beta_3$  & $\gamma_0$ & $\gamma_1$ &$\gamma_2$ & $\gamma_3$ \\
  & Simulated  &0.303 & 0.173 & 1.785 & -0.090 & 66.275 & -22.496 & -15.701 & -9.717\\
  & lower limit &-0.903 & -0.546 & 1.024 & -0.156 & 22.557 & -46.161 & -33.822 & -14.432 \\
  & upper limit & 1.349 & 1.004 & 2.556 & -0.004 & 97.254 & -7.479 & -5.565 & -2.986 \\
\hline
& \textbf{Full data} &-0.525& 0.499 &2.326 & -0.05 & 62.847 &-19.147 & -8.665 & -10.619\\
\hline
    \end{tabular}
\caption{Parameters estimators and confidence intervals using 20\%, 30\%, 40\% and 50\% of lost data}\label{pc2}
\end{table}
Table \ref{pc2} shows the confidence intervals at $95\%$ for the parameters of the fixed effects of the model and the mean of the estimates. Also, this table shows that the confidence intervals contain the estimated parameters of the complete data shown in Table \ref{pc1}. It is noteworthy that as the percentage of random loss increases, the size of the interval increases. In addition, as the percentage of missing information decreases, the average of the simulated coefficients approaches the values of the parameters observed in Table \ref{pc1} which are observed in the lower part of Table \ref{pc2} (full data).
The average success rate of the algorithm, for each of the loss percentages, is shown in Table \ref{two}. The success rate of the algorithm is on average $72\%$, the confidence intervals for the fixed effects shown in Table \ref{pc2} contain the parameters of the complete model evidencing the benefits of the algorithm proposed for the estimation of the missing information.

\begin{table}[ht]
\centering
\begin{tabular}{|r|rrrr|}
   \hline
Missing&$20\%$ & $30\%$ & $40\%$ & $50\%$ \\
\hline
Success&$83.82\%$ & $77.59\%$ & $75.96\%$ & $72.38\%$ \\
    \hline
\end{tabular}
\caption{Algorithm success with Zero-inflated Poisson response.}\label{two}
\end{table}
\section{Conclusions}
By proposing the use of the EM algorithm and Bartlett's ANCOVA method for the estimation and imputation of missing information in response variables of the zero-inflated Poisson type, a methodology is achieved that is useful in situations in which the covariates are fully observed, allowing the estimation of the parameters simply.

The proposed methodology estimates and imputes the missing values using the EM algorithm in two steps each time. In the initial step, it proposes specific weights for the values of the response variable, taking into account whether the data is observed or missing, and performs a weighted regression. In the second step, the missing information is imputed and the model parameters are re-estimated. In addition, by doing it sequentially through time, it allows working with individuals who leave the study at some time $t>1$ and allows the reincorporation of the individual, which helps to fill in very common information in clinical or agricultural studies.

{In the simulation exercise, the good performance of the methodology is evident to estimate the missing data and allow working with complete information, especially when there are not so many experimental units in each treatment. It also allows the estimation of the parameters to be constructed and reduces the bias concerning models that eliminate or fill the missing data with the mode.}

The application of the proposed methodology was carried out, for the inflated zero Poisson, to real data and it was observed that estimated values had average hits of $72\%$ concerning the original data. Also, the confidence intervals contain the estimated parameters of the model made with the data from the maize study.

For future work, it is recommended to extend the methodology to cases in which the response variable follows another type of discrete or continuous distribution. For example, zero-inflated beta in the continuous case and zero-inflated geometric in the discrete case. Make use of generalized mixed-effects linear models or generalized estimating equations for the estimation of the stepwise algorithm and the weights of missing data. In addition, determines the influence that can be presented in the estimation of the model, the possible imputation of atypical data, as well as determine the efficiency of the model according to the sample size.

\section*{Supplementary files}
\begin{enumerate}
    \item\label{SF1} Additional figures from the simulation exercise along with the R code of the simulation.
    \item Application when data loss was $20\%$, with step-by-step development of tables, estimated weights and imputed data.
    
\end{enumerate}



\appendix \label{ApenAA}

\section{}
\subsection{Time 2}
In the estimation of the model parameters for the second time, the response variable of the time $t=1$, $y_{i1}$ is used as a covariate associated with the model using the $EM$ algorithm with weights. During the imputation of the missing values for the time $t=1$, all the entries of the response variable may be zero, in which case imputing the data for the time $t=1$ does not improve the explanation of the variability of the response variable. Therefore, if this case happens, this variable should not be considered and the link functions for the parameters $\lambda_{i2}$ and $\pi_{i2}$ should be defined as:
\begin{align}
     \ln[\lambda_{i2}]=\pmb{x}_{i2}'\pmb{\beta}  & \qquad \text{if} \ y_{i2} > 0 \nonumber \\
\ln\left[\dfrac{\pi_{i2}}{1-\pi_{i2}}\right]=\pmb{z}_{i2}'\pmb{\gamma} & \qquad \text{if} \ y_{i2} = 0 \label{A26}
\end{align}
The estimation of the model parameters and the imputation of the data is carried out in a similar way as proposed for the time $t=1$.
However, if the inputs of $y_{i1}$ are not all zero, which is the most probable situation, the estimation of the vector of maximum plausible parameters in time $t=2$ must be carried out in the same way as in Step 1 of time $ t=1$, with the use of the $EM$ algorithm taking into account $y_{i1}$ as a covariate associated in the model, that is, the link functions for the parameters are as in \eqref{A26}. These can be written in matrix form as:
\begin{align*}
    E[ g(\pmb{y}_{2})]= \begin{pmatrix}
\pmb{X}_{2} &\pmb{Z}_{2} \end{pmatrix}
\begin{pmatrix}
\pmb{\beta} \\
\pmb{\gamma} \\
\end{pmatrix}
+ \pmb{e}
\end{align*}
where $\pmb{X}_{2}$ and $\pmb{Z}_{2}$ are the matrices of size $n\times (p+1)$ of covariates related to $\pmb{y}_ {2}$, which is the response vector at time $t=2$. There are $p$ columns containing the independent variables that explain the behavior of the dependent variable and the last column is the response vector at time $t= $1 using the previously imputed data. The $\log$-$\text{likelihood}$ is defined as:

\begin{align}\label{28}
  \ln \ell_2 &=\sum_{y_{i2}=1} \ln[\pi_{i2}+(1-\pi_{i2}) \exp{(-\lambda_{i2}})] \nonumber \\
& \quad +\sum_{y_{i2}> 0} \left\{ \ln(1-\pi_{i2})-\lambda_{i2}+y_{i2}\ln(\lambda_{i2}) -\ln(y_{i2}!) \right\}
\end{align}
and substituting \eqref{A26}:
\begin{align*}
\ln \ell_2=& \sum_{y_{i2}>0}\left\{
-\ln[1+\exp{(\pmb{z}_{i2}'\pmb{\gamma})}]-\exp{(\pmb{x}_{i2}'\pmb{\beta} )} \right\}+\sum_{y_{i2}>0}\left\{y_{i2}\left(\pmb{x}_{i2}'\pmb{\beta} \right)-\ln(y_{i2}!)
\right\}\\
&+ \sum_{y_{i2}=0}\ln\left[
\dfrac{\exp{(\pmb{z}_{i2}'\pmb{\gamma})}}{1+\exp{(\pmb{z}_{i2}'\pmb{\gamma}) }}+\dfrac{\exp{(-\exp{(\pmb{x}_{i2}'\pmb{\beta})})}}{1+\exp{(\pmb{z}_{ i2}'\pmb{\gamma})}}\right]\\
&=\sum_{i=1}^{n}
D_{i2}\ln\left[\exp{(\pmb{z}_{i2}'\pmb{\gamma})}
+\exp{(-\exp{(\pmb{x}_{i2}'\pmb{\beta})})} \right] \\
& \quad + \sum_{i=1}^{n}
(1-D_{i2})\left[y_{i2}\left(\pmb{x}_{i2}'\pmb{\beta} \right) -\exp{(\pmb{x}_{i2 }'\pmb{\beta})}-\ln (y_{i2}!)\right] \\
& \quad -
  \sum_{i=1}^{n}\ln[1+\exp{(\pmb{z}_{i2}'\pmb{\gamma})}]
\end{align*}
If the indicator function is $D_{i2}=0$, it is obtained that:
\begin{align*}
  D_{i2}\ln\left[\exp{(\pmb{z}_{i2}'\pmb{\gamma})}
+\exp{(-\exp{(\pmb{x}_{i2}'\pmb{\beta})})} \right]=0
\end{align*}
On the other hand, if the indicator function is $D_{i2}=1$ then the model seeks to estimate the zero state, and therefore, the covariates associated with the Poisson model are not taken into account \cite{lambert1992zero}. Then, it is obtained that:
\begin{equation*}
    \begin{split}
\ln \ell_2&= \ell(D_{i2},\pmb{\gamma},y_{i2}) + \ell( D_{i2},\pmb{\beta}, y_{i2})
    \end{split}
\end{equation*}
where
\begin{align*}
\ell(D_{i2},\pmb{\gamma},y_{i2})&=\sum_{i=1}^{n}\left\{ D_{i2}(\pmb{z}_{i2}'\pmb{\gamma})
-\ln[1+\exp{(\pmb{z}_{i2}'\pmb{\gamma})}]\right\}\\
\ell( D_{i2},\pmb{\beta}, y_{i2})&=
\sum_{i=1}^{n}(1-D_{i2})\left[y_{i2}(\pmb{x}_{i2}'\pmb{\beta})-
\exp{(\pmb{x}_{i2}'\pmb{\beta})}-\ln(y_{i2}!)
\right]
\end{align*}
Therefore, the likelihood of $\ell(D_{i2},\pmb{\gamma},y_{i2})$ and $\ell( D_{i2},\pmb{\beta},y_{i2} )$ can be separately maximized using the $EM$ algorithm iteratively, toggling between the estimate of $D_{i2}^{(0)}$ given the conditional expectation under some initial parameters $\pmb{\gamma} ^{(0)}$ and $\pmb{\beta}^{(0)}$ initially estimated. Subsequently, with $D_{i2}^{(0)}$ estimated, the function is maximized to obtain the estimate of $\pmb{\gamma}^{(1)}$ and $\pmb{\beta}^{( 1)}$, and then iteratively until the $m$th iteration where convergence is achieved.

\subsubsection{Step 1}
The estimated values for $\pmb{\beta}$ and $\pmb{\gamma}$ are found using the $EM$ algorithm for the data at time $t=2$.
\paragraph{Step E.} Estimate $D_{i2}$ given the conditional mean $D_{i2}^{(m)}$ under the initial estimates of $ \pmb{\beta}^{(m)}$ and $\pmb{ \gamma}^{(m)}$, that is:

\begin{align*}
     E[D_{i2}^{(m)}\vert y_{i2},\pmb{\beta}^{(m)},\pmb{\gamma}^{(m)}]
&=P[D_{i2}^{(m)}=1\vert y_{i2},\pmb{\beta}^{(m)},\pmb{\gamma}^{(m)}]
\end{align*}
Then, by Bayes' theorem, it is obtained that:
{\footnotesize
\begin{align*}
&P\left[D_{i2}^{(m)}=1\vert y_{i2},\pmb{\gamma}^{(m)}, \pmb{\beta}^{(m)}\right ]= \\
& \small{
\dfrac{
P\left[y_{i2}=0,\pmb{\gamma}^{(m)}, \pmb{\beta}^{(m)}\vert D_{i2}^{(m)}=1 \right] 
P\left[D_{i2}^{(m)}=1 \right] }
{
P\left[y_{i2}=0,\pmb{\gamma}^{(m)}, \pmb{\beta}^{(m)}\vert D_{i2}^{(m)}=1 \right]
P\left[D_{i2}^{(m)}=1 \right]+
P\left[y_{i2}> 0,\pmb{\gamma}^{(m)}, \pmb{\beta}^{(m)}\vert D_{i2}^{(m)}= 0\right]
P\left[D_{i2}^{(m)}=0 \right]
}}
\end{align*}
}
By the definition of the indicator function, it is clear that $P[D_{i2}^{(m)}=1]=\pi_{i2}^{(m)}$ and similarly $P[D_{i2} ^{(m)}=0]=1-\pi_{i2}^{(m)}$, then it is obtained that:

\begin{align*}
E\left[D_{i2}^{(m)}\vert y_{i2},\pmb{\gamma}^{(m)},\pmb{\beta}^{(m)} \right] & = \begin{cases}
\dfrac{1}{1+ \exp{[-\exp{(\pmb{x}_{i2}'\pmb{\beta}^{(m)} )} -\pmb{z}_{i2 }'\pmb{\gamma}^{(m)} ]} } & \ \qquad \text{if} \ y_{i2}=0 \\
0 & \ \qquad \text{if} \ y_{i2}> 0\\
\end{cases}
\end{align*}
\paragraph{Step M.} $\pmb{\gamma}^{(m)}$ is obtained by maximizing $\ell(\pmb{\gamma}^{(m)}, y_{i2}, D_ {i2}^{(m)})$, analogous to time $t=1$. Therefore, the equation to be solved is:
\begin{equation*}
  \sum_{i=1}^{n} \pmb{z}_{i2}' \left( D_{i2}^{(m)}-\pi_{i2}^{(m)} \right)w_ {i2(k)}^{(m)}=0
\end{equation*}
where $w_{i2(k)}^{(m)}$ is the weight corresponding to the $i$th observation in the $m$th iteration {assuming that the value it will take is $k$} in time 2. These values are specified in the same way as at time 1, that is:
{
\begin{align*}
w_{i2(k)}^{(m)} &=  \begin{cases}
1 &\text{if the data $y_{i2}$ is observed and $k=y_{i1}$}\\
0 &\text{if the data $y_{i2}$ is observed and $k\neq y_{i1}$}\\
\pi_{i2}^{(m)}+(1-\pi_{i2}^{(m)})\exp{(-\lambda_{i2}^{(m)})}&\text{if the data $y_{i2}$ is missing and} \  k= 0\\
\dfrac{(1-\pi_{i2}^{(m)})\exp{(-\lambda_{i2}^{(m)})} (\lambda_{i2}^{(m)})^{k}}{(k)!}&\text{if the data $y_{i2}$ is missing and} \  k> 0\\
\end{cases}
\end{align*}
}
To estimate $\pmb{\gamma}$, the Scoring-Fisher method is used, that is
\begin{align}\label{30}
   \pmb{\gamma}^{(m+1)}&=  \pmb{\gamma}^{(m)}+ [\pmb{\Im}_{z}^{(m)}]^{-1}\pmb{U}_{z}^{(m)}
 \end{align}
with
\begin{align*}
\pmb{\Im}_{z}^{(m)}&=E\left[- \dfrac{\partial^{2} \ell(\pmb{\gamma}^{(m)}\vert y_{i2},D_{i2}^{(m)})}
{\partial \pmb{\gamma}' \partial \pmb{\gamma}} \right]=\sum_{i=1}^{n}\pmb{z}_{i2}' \pi_{i2}^{(m)}(1-\pi_{i2}^{(m)})w_{i2(k)}^{(m)}\pmb{ z}_{i2}\\
&=\pmb{Z}_{2}'\pmb{M}_{2z}^{(m)}\pmb{W}_{2}^{(m)}\pmb{Z}_{2}
\end{align*}
where $\pmb{M}_{2z}^{(m)}=diag(\pi_{i2}^{(m)}(1-\pi_{i2}^{(m)}))$ and $ \pmb{W}_{2}^{(m)}=diag(w_{i2(k)}^{(m)})$ are matrices of size $n\times n$ and $Z_{2}$ is the matrix of covariates associated with the zero-inflated model at time 2 of size $n\times (p+1)$. Furthermore, it is known that:
\begin{equation}\label{35}
      \pmb{U}_{z}^{(m)}= \sum_{i=1}^{n}
\pmb{z}_{i2}'\left(D_{i2}^{(m)}-\pi_{i2}^{(m)} \right)\pmb{w}_{i2(k)} ^{(m)}=\pmb{Z}_{2}'\pmb{M}_{2z}^{(m)}\pmb{W}_{2}^{(m)}\pmb{v}_{z2 }^{(m)}
\end{equation}
where
{
\begin{align}
  \pmb{v}_{z2}^{(m)}&=\Bigg( \dfrac{D_{12}^{(m)}-\pi_{12}^{(m)}  }{\pi_{12}^{(m)}(1-\pi_{12}^{(m)}) }  \ldots, \dfrac{D_{i2}^{(m)}-\pi_{i2}^{(m)} }{\pi_{i2}^{(m)}(1-\pi_{i2}^{(m)}) } ,
  \ldots,   \dfrac{D_{n2}^{(m)}-\pi_{n2}^{(m)}}{\pi_{n2}^{(m)}(1-\pi_{n2}^{(m)}) } \Bigg)'\nonumber
\end{align}
}
Replacing \eqref{35} in \eqref{30}, it is obtained that:
\begin{align*}
\pmb{\gamma}^{(m+1)}= \pmb{\gamma}^{(m)}
+ \left[ \pmb{Z}_{2}'\pmb{M}_{2z}^{(m)} \pmb{W}_{2}^{(m)}\pmb{Z}_ {2}\right] ^{-1} 
\pmb{Z}_{2}'\pmb{M}_{2z}^{(m)}\pmb{W}_{2}^{(m)}\pmb{v}_{z2}^ {(m)}
\end{align*}

Similarly, it is derived from the $\log$-$\text{likelihood}$ for $\pmb{\beta}$ as follows:
\begin{align*}
    \begin{split}
\pmb{\beta}^{(m+1)}=\pmb{\beta}^{(m)}
+ \left[
\pmb{X}_{2}' \pmb{M}_{x2}^{(m)}\pmb{W}_{2}^{(m)}\pmb{X}_{2} \right]^{-1}  \pmb{X}_{2}'\pmb{M}_{x2}^{(m)}\pmb{W}_{2}^{(m)}\pmb{v}_{x2}^{(m)}
    \end{split}
\end{align*}
where $\pmb{M}_{x2}^{(m)}=diag(\lambda_{i2}^{(m)})$ is a matrix of size $n \times n $, $\pmb{X }_{2}$ is the covariate matrix of size $n \times p$, and
{
\small
\begin{align*}
\pmb{v}_{x2}^{(m)}&= \left( \dfrac{(y_{12}^{(m)}) \left[ y_{12}-\lambda_{12}^{(m)} \right]}{\lambda_{12}^{(m)}}, \ldots,  \dfrac{(y_{i2}^{(m)}) \left[ y_{i2}-\lambda_{i2}^{(m)} \right]}{\lambda_{i2}^{(m)}} ,\ldots,\dfrac{(y_{n2}^{(m)}) \left[ y_{n2}-\lambda_{n2}^{(m)} \right]}{\lambda_{n2}^{(m)}} \right)'
\end{align*}
}
is a vector of size $n \times 1$. After having the estimators, the missing data is imputed at time 2 in the following section.
\subsubsection{Step 2}
\paragraph{Step E.} The same procedure given in step 2 is carried out at time $t=1$, to obtain the estimates of the missing data at time 2. The model to be taken into account is:
\begin{align*}
    E[g(\pmb{y}_{2})\vert \pmb{X}_{2},\pmb{Z}_{2}]=
\begin{pmatrix}
\pmb{X}_{2} \\
\pmb{0}
\end{pmatrix}
  \pmb{\beta}
+
\begin{pmatrix}
\pmb{0} \\
\pmb{Z}_{2}
\end{pmatrix}
  \pmb{\gamma}
+
\begin{pmatrix}
\pmb{A}_{2} \\
\pmb{0}
\end{pmatrix}
  \pmb{\alpha}
+
\begin{pmatrix}
\pmb{0} \\
\pmb{B}_{2}
\end{pmatrix}
  \pmb{\tau}
\end{align*}
The parameters $\pmb{\alpha}$ and $\pmb{\tau}$ are the estimated coefficients for the covariates of the missing values and $\pmb{A}_{2}$ and $ \pmb{B}_ {2}$ correspond to the second-time missing value covariate matrices for the Poisson and zero-inflated models, respectively. By making the same deduction as step 2 of time 1, it is obtained that:
\begin{align*}
\widehat{\pmb{\alpha}}&=
\pmb{X}_{2(miss)}\widehat{\pmb{\beta}} \\
\widehat{\pmb{\tau}}&=
\pmb{Z}_{2(miss)}\widehat{\pmb{\gamma}}
\end{align*}
The other characteristics are deduced in the same way as in step 2 of time 1. That is, taking into account that the distribution of the response variable is Poisson with an excess of zeros, the models to be estimated are:
\begin{equation*}
    \begin{cases}
\widehat{\alpha}^{(m)}_{i}=\pmb{x}_{i2(miss)}'\widehat{\pmb{\beta}}^{(m)}=\ln[\widehat{\lambda}_{i2(miss)}^{(m)}]  & \text{if} \ y_{i2} > 0   \\
\widehat{\tau}^{(m)}_{i}=\pmb{z}_{i2(miss)}'\widehat{\pmb{\gamma}}^{(m)}=\ln\left[\dfrac{\widehat{\pi}_{i2(miss)}^{(m)}}{1-\widehat{\pi}_{i2(miss)}^{(m)}}\right] &\text{if} \ y_{i2}= 0
    \end{cases}
\end{equation*}
Therefore,
\begin{equation*}
    \begin{cases}
\widehat{\lambda}_{i2(miss)}^{(m)} =\exp{(\pmb{x}_{i2(miss)}'\widehat{\pmb{\beta}}^{ (m)} ) }&\text{if } \ y_{i2} > 0 \\
     \widehat{\pi}_{i2(miss)}^{(m)}=\dfrac{\exp{(\pmb{x}_{i2(miss)}'\widehat{\pmb{\gamma} }^{(m)} })}{1+\exp{(\pmb{z}_{i2(miss)}'\widehat{\pmb{\gamma}}^{(m)} ) }} &\text{if } \ y_{i2} = 0\\
    \end{cases}
\end{equation*}
where $\hat{\lambda}_{i2}^{(m)}$ and $\hat{\pi}_{i2}^{(m)}$ refer to the value of the mean corresponding to the individual $i$ at time 2.
The selection criteria are the same as in step 2 of time 1.

\paragraph{Step M.} Once the observations are imputed, maximization starting from the complete data set is carried out with the Fisher-Scoring method, with the initial estimator given in step 1 of time $t=2$. Iterate between steps E and M until convergence is achieved. The expressions are:
\begin{align*}
E\left[D_{i2}^{(m)}\vert y_{i2},\pmb{\gamma}^{(m)},\pmb{\beta}^{(m)} \right]& = \begin{cases}
\dfrac{1}{1+ \exp{(-\pmb{z}_{i2}'\pmb{\gamma}^{(m)}
-\exp{(\pmb{x}_{i1}'\pmb{\beta}^{(m)})})} } &\text{if } \ y_{i2}=0 \\
0&\text{if } \ y_{i2}> 0\end{cases}
\end{align*}
If the following is defined:
{
\begin{align*}
\pmb{v}_{z2}^{(m)}&=\Bigg( \dfrac{D_{12}^{(m)}-\pi_{12}^{(m)}  }{\pi_{12}^{(m)}(1-\pi_{12}^{(m)}) }  \ldots, \dfrac{D_{i2}^{(m)}-\pi_{i2}^{(m)} }{\pi_{i2}^{(m)}(1-\pi_{i2}^{(m)}) } ,
  \ldots,   \dfrac{D_{n2}^{(m)}-\pi_{n2}^{(m)}}{\pi_{n2}^{(m)}(1-\pi_{n2}^{(m)}) } \Bigg)' \\
\pmb{M}_{2z}^{(m)}&=diag(
\pi_{i2}^{(m)}(1-\pi_{i2}^{(m)}))
\end{align*}
}
where $\pmb{M}_{2z}$ is a matrix of size $n\times n$, and $\pmb{v}_{z2}$ is a vector of $n\times 1$. Then, the Fisher-Scoring algorithm for $\pmb{\gamma}$ is:
\begin{align*}
\pmb{\gamma}^{(m+1)}
=\pmb{\gamma}^{(m)}+ \left[
\pmb{Z}_{2}'\pmb{M}_{2z}^{(m)} \pmb{Z}_{2}\right]^{-1}  \left(
\pmb{Z}_{2}'\pmb{M}_{2z}^{(m)}\pmb{v}_{z2}^{(m)} \right)
\end{align*}
where $\pmb{Z}_{2}$ is a matrix of covariates of $n \times (p+1)$. Similarly, for $\pmb{\beta}$ the expressions are:
{
\small
\begin{align*}
\pmb{M}_{x2}^{(m)}&=diag(\lambda_{i2}^{(m)})\\
\pmb{v}_{x2}^{(m)}&= \left( \dfrac{(1-D_{12}^{(m)}) \left[ y_{12}-\lambda_{12}^{(m)} \right]}{\lambda_{12}^{(m)}}, \ldots,  \dfrac{(1-D_{i2}^{(m)}) \left[ y_{i2}-\lambda_{i2}^{(m)} \right]}{\lambda_{i2}^{(m)}} ,\ldots,\dfrac{(1-D_{n2}^{(m)}) \left[ y_{n2}-\lambda_{n2}^{(m)} \right]}{\lambda_{n2}^{(m)}} \right)'
\end{align*}
}
where $\pmb{M}_{x2}$ is of size $n \times n$, and $\pmb{v}_{x2}$ of size $n \times 1$. So:
\begin{align*}
    \begin{split}
\pmb{\beta}^{(m+1)}=\pmb{\beta}^{(m)}+ \left[
\pmb{X}_{2}' \pmb{M}_{x2}^{(m)}\pmb{X}_{2}\right]^{-1}  \pmb{X}_{2}'\pmb{M}_{x2}^{(m)}\pmb{v}_{x2}^{(m)}
    \end{split}
\end{align*}
where $\pmb{X}_{2}$ is a covariate matrix of $n \times (p+1)$.
\appendix \label{ApenAB}
\section{}
\subsection{Time $t$}
\subsubsection{Step 1}
$C_{t-1}$ is defined as the eigenvector associated with the responses at times $t-1$. The estimation of the vector of maximum plausible parameters at time $t$ is carried out in the same way as at time 1 using the algorithm $EM$. The binding functions for the parameters are
\begin{equation}
    \begin{split}
    \ln[\lambda_{it}]=\pmb{x}_{it}'\pmb{\beta} & \qquad \text{if} \ y_{it} > 0  \\
\ln\left[\dfrac{\pi_{it}}{1-\pi_{it}}\right]=\pmb{z}_{it}'\pmb{\gamma} &\qquad \text{if} \ y_{it} = 0
\end{split}
\end{equation}
where $\pmb{x}'_{it}$ and $\pmb{z}'_{it}$ are vectors of size $p+1$, where $p$ columns refer to the covariates that explain the behavior of the variable at time $t$ and the last column is the first eigenvector of the principal component analysis performed with the values of $y_{i1},y_{i2},\ldots, y_{i(t -1)}$. The weights for the time $t$ are defined as:
{
\begin{align*}
w_{it(k)}^{(m)} &=  \begin{cases}
1 &\text{if the data $y_{it}$ is observed and $k=y_{it}$}\\
0 &\text{if the data $y_{it}$ is observed and $k\neq y_{it}$}\\
\pi_{it}^{(m)}+(1-\pi_{it}^{(m)})\exp{(-\lambda_{it}^{(m)})}&\text{if the data $y_{it}$ is missing and} \  k= 0\\
\dfrac{(1-\pi_{it}^{(m)})\exp{(-\lambda_{it}^{(m)})} (\lambda_{it}^{(m)})^{k}}{(k)!}&\text{if the data $y_{it}$ is missing and} \  k> 0\\
\end{cases}
\end{align*}
}
and the matrices for the estimation of the model parameters at time $t$ are derived the same as at step 1 from time 1. That is, $\pmb{M}_{tz}^{(m)}= diag(
\pi_{it}^{(m)}(1-\pi_{it}^{(m)})$ and $W_{t}^{(m)}=diag(w_{it(k)}^ {(m)})$ of sizes $n \times n$, and the vector
{
\begin{align*}
\pmb{v}_{zt}^{(m)}&=\Bigg( \dfrac{D_{1t}^{(m)}-\pi_{1t}^{(m)}  }{\pi_{1t}^{(m)}(1-\pi_{1t}^{(m)}) }  \ldots, \dfrac{D_{it}^{(m)}-\pi_{it}^{(m)} }{\pi_{it}^{(m)}(1-\pi_{it}^{(m)}) } ,\ldots,   \dfrac{D_{nt}^{(m)}-\pi_{nt}^{(m)}}{\pi_{nt}^{(m)}(1-\pi_{nt}^{(m)}) } \Bigg)'
\end{align*}
}
of size $n \times 1$. Therefore, after applying Scoring Fisher, it is obtained that:
\begin{align*}
\pmb{\gamma}^{(m+1)} =
\pmb{\gamma}^{(m)}+ \left[
\pmb{Z}_{t}'\pmb{M}_{tz}^{(m)} \pmb{W}_{t}^{(m)}\pmb{Z}_{t} \right]^{-1}    
\pmb{Z}_{t}'\pmb{M}_{tz}^{(m)}\pmb{W}_{t}^{(m)}\pmb{v}_{zt}^{(m)} \end{align*}
with $\pmb{Z}_{t}$ of size $n \times (p+1)$. Similarly, let $\pmb{M}_{xt}^{(m)}=diag(\lambda_{it}^{(m)})$ be the matrix  of size $n \times n$, and the vector
{
\small
\begin{align*}
\pmb{v}_{xt}^{(m)}&= \left( \dfrac{(1-D_{1t}^{(m)}) \left[ y_{1t}-\lambda_{1t}^{(m)} \right]}{\lambda_{1t}^{(m)}}, \ldots,  \dfrac{(1-D_{it}^{(m)}) \left[ y_{it}-\lambda_{it}^{(m)} \right]}{\lambda_{it}^{(m)}} ,\ldots,\dfrac{(1-D_{nt}^{(m)}) \left[ y_{nt}-\lambda_{nt}^{(m)} \right]}{\lambda_{nt}^{(m)}} \right)'
\end{align*}
}
It is obtained that
\begin{align*}
    \begin{split}
\pmb{\beta}^{(m+1)}
= \pmb{\beta}^{(m)}+ \left[
\pmb{X}_{t}' \pmb{M}_{xt}^{(m)}\pmb{W}_{t}^{(m)}\pmb{X}_{t}  \right]^{-1}
\pmb{X}_{t}'\pmb{M}_{xt}^{(m)}\pmb{W}_{t}^{(m)}\pmb{v}_{xt}^{(m)} \end{split}
\end{align*}
where $\pmb{X}_{t}$ is a covariate matrix of size $n \times (p+1)$. For the estimation and imputation of missing data in time $t$, the EM algorithm is carried out for the data in time $t$ taking into account as estimator of the parameters obtained in this step.
\subsubsection{Step 2}
\paragraph{Step E.} The same procedure given in step 2 of time $t=1$ is performed. To obtain the estimates of the missing data at time 3, the model is taken into account:
\begin{align*}
    E[g(\pmb{y}_{t})\vert \pmb{X}_{t},\pmb{Z}_{t}]=
\begin{pmatrix}
\pmb{X}_{t} \\
\pmb{0}
\end{pmatrix}
  \pmb{\beta}
+
\begin{pmatrix}
\pmb{0} \\
\pmb{Z}_{t}
\end{pmatrix}
  \pmb{\gamma}
+
\begin{pmatrix}
\pmb{A}_{t} \\
\pmb{0}
\end{pmatrix}
  \pmb{\alpha}
+
\begin{pmatrix}
\pmb{0} \\
\pmb{B}_{t}
\end{pmatrix}
  \pmb{\tau}
\end{align*}
where $\pmb{X}_{t}$ and $\pmb{Z}_{t}$ are matrices of covariates at time $t$ of size $n \times (p+1)$.
The coefficients $\pmb{\beta}$ and $\pmb{\gamma}$ are the parameters associated with the regression of the covariates. The parameters $\pmb{\alpha}$ and $\pmb{\tau}$ are the estimated coefficients for the covariates of the missing values, and $\pmb{A}_{t}$ and $ \pmb{B}_{t}$ correspond to the matrices that of the covariates of the missing value of the time $t$. When performing the same deduction as step 2 of the time $t=1$, the following is obtained:
\begin{align*}
\widehat{\pmb{\alpha}}&=
\pmb{X}_{t(miss)}\widehat{\pmb{\beta}} \\
\widehat{\pmb{\tau}}&=
\pmb{Z}_{t(miss)}\widehat{\pmb{\gamma}}
\end{align*}

\begin{center}
$\widehat{y}_{it(miss)} = \begin{cases}
0&\text{if } \   \widehat{\pi}_{it(miss)}^{(m)} >p_{o} \\
  \widehat{\lambda}_{it(miss)}^{(m)}&\text{if } \     \widehat{\pi}_{it(miss)}^{(m)}\leq p_{o}\\
\end{cases} $
\end{center}
\paragraph{Step M.} Once the observations have been imputed, the maximization starts from the complete data set, considering the Fisher-Scoring method with the initial estimator given in step 1. Steps E and M are repeated until achieving convergence. The expressions are:
\begin{align*}
E\left[D_{it}^{(m)}\vert y_{it},\pmb{\beta}^{(m)},\pmb{\gamma}^{(m)}\right] & = \begin{cases}
\dfrac{1}{1+ \exp{(-\pmb{z}_{it}'\pmb{\gamma}^{(m)}
-\exp{(\pmb{x}_{it}'\pmb{\beta}^{(m)})})} } &\text{if } \ y_{it}=0 \\
0&\text{if } \ y_{it}> 0\end{cases}
\end{align*}
\begin{align*}
\pmb{\gamma}^{(m+1)}
= \pmb{\gamma}^{(m)}+ \left[
\pmb{Z}_{t}'\pmb{M}_{tz}^{(m)} \pmb{Z}_{t} \right]^{-1}  \pmb{Z}_ {t}'\pmb{M}_{tz}^{(m)}\pmb{v}_{zt}^{(m)} \end{align*}
with $\pmb{Z}_{t}$ the covariate matrix of size $n \times (p+t-1)$. Similarly, for $\pmb{\beta}$ it is obtained that:
\begin{align*}
     \begin{split}
\pmb{\beta}^{(m+1)}
= \pmb{\beta}^{(m)}+ \left[
\pmb{X}_{t}' \pmb{M}_{xt}^{(m)}\pmb{X}_{t} \right]^{-1} 
\pmb{X}_{t}'\pmb{M}_{xt}^{(m)}\pmb{v}_{xt}^{(m)} \end{split}
\end{align*}
with $\pmb{X}_{t}$ of size $n \times (p+1)$.

\appendix\label{ApenAC}
\section{}
Table \ref{p1} shows $20\%$ of missing data simulating the missing data from Table \ref{origin1}.
\begin{table}[ht]
\centering
\begin{tabular}{rrrrrrrrrrr}
   \hline
  & We1 & We2 & We3 & We4 & We5 & We6 & We7 & We8 & We9 & Treat \\
   \hline
1 & 0 & 0 & 0 & 0 &  & 0 &  & 0 & 0 \\ 
  2 & 0 & 0 &  & 0 & 0 & 0 & 0 &  & 0 \\ 
  3 & 0 &  & 0 & 0 & 0 & 0 & 0 & 0 & 0 \\ 
  4 &  & 0 &  &  & 0 &  & 0 & 0 & 1 \\ 
  5 &  &  & 0 & 0 &  & 0 & 0 & 0 & 1 \\ 
  6 &  & 0 & 0 & 0 &  & 0 & 0 &  & 1 \\ 
  7 & 0 & 0 & 0 & 0 & 0 & 1 &  & 1 & 2 \\ 
  8 & 0 & 0 & 0 & 0 & 0 &  & 0 & 1 & 3 \\ 
  9 & 0 & 0 &  & 0 & 0 & 0 & 0 & 0 & 0 \\ 
  10 & 0 & 0 & 0 & 0 & 0 & 0 & 0 & 0 & 0 \\ 
  11 & 0 & 0 & 0 & 0 & 0 & 0 & 0 & 0 & 0 \\ 
  12 & 0 & 0 & 0 & 0 & 0 & 0 & 0 & 1 & 0 \\ 
  13 & 0 & 0 & 0 & 1 & 1 & 1 & 0 & 1 & 0 \\ 
  14 & 0 & 0 & 0 & 0 & 1 &  & 1 & 1 & 0 \\ 
  15 &  & 0 &  & 0 & 1 & 2 &  &  & 1 \\ 
  16 & 0 & 0 & 0 & 0 & 2 & 4 &  & 2 & 3 \\ 
  17 & 0 & 0 & 0 &  & 1 & 4 & 3 &  & 2 \\ 
  18 & 0 & 0 & 0 & 0 & 1 & 5 & 4 &  & 3 \\ 
  19 & 0 & 0 & 0 & 0 & 0 & 5 & 4 & 2 & 3 \\ 
  20 & 0 & 0 & 0 &  & 0 & 5 & 5 & 2 & 4 \\ 
  21 & 0 & 0 & 0 & 0 & 0 & 4 &  & 3 & 4 \\ 
  22 & 0 & 0 & 0 & 0 & 0 & 8 & 6 & 3 & 6 \\ 
  23 &  & 0 &  & 0 & 0 &  & 7 & 4 & 4 \\ 
  24 &  &  & 0 & 0 & 0 & 9 &  &  & 4 \\
    \hline
\end{tabular}
\caption{Corn data: $20\%$ missing data.}\label{p1}
\end{table}
\paragraph{Step 1.}
For the imputation of the data in Table \ref{p1}, the following models are proposed at each time:
\begin{enumerate}
    \item For times 1, 2, 3, 4, and 5, the model to be estimated is proposed as::
    \begin{align*}
\ln{\left(\dfrac{\pi_{it(miss)}}{1-\pi_{it(miss)}}\right)}= & \gamma_{0t}+\gamma_{1t}x_{i1}+\gamma_{2t}x_{i2}\\
   \ln(\lambda_{it(miss)})= & \beta_{0t}+\beta_{1t}x_{i1}+\beta_{2t}x_{i2}, \qquad t=1,2,3,4,5 \\
\end{align*}
where $\pi_{it(miss)}$ is the probability that the missing data at time $t$ belongs to the zero model, $\lambda_{it(miss)}$ is the mean of the Poisson model and $T_{it}$ is the covariate that refers to the treatment of the $i$th observation. For times $t=1,2,3,4,5$, treatment is taken as the only covariate, taking into account that the values of times: 1, 2 and 3 are all zero and that time 4 only has a non-zero observation. Parameters estimation for $t=2,3,4,5$, with any of the above times, generates model estimation problems.  The parameters estimated for $\gamma_{0t}$, $\gamma_{1t}$, $\beta_{0t}$ and $\beta_{1t}$ for times 1, 2, 3, 4 and 5 are seen in Table \ref{table12345}.
\item At time 6, the model to be estimated is proposed as:
\begin{align*}
\ln{\left(\dfrac{\pi_{i6(miss)}}{1-\pi_{i6(miss)}}\right)}= & \gamma_{06}+\gamma_{16}T_{i }+ \gamma_{26}t_{5}\\
   \ln(\lambda_{i6(miss)})= & \beta_{06}+\beta_{16}T_{i}+\beta_{26}t_{5}
\end{align*}
where $t_{5}$ are the values of the response variable at time 5.
\item At time 7, the responses of times 4, 5, and 6 are taken, and a principal component analysis is performed to obtain a new covariate that retains the maximum variability contained in the data. Analogously, at time 8 the responses from 4 to 7 are taken and for time 9, the responses from time 4 to 8 are taken, and a principal component analysis is performed to obtain a new covariate that retains the maximum variability contained in the data.  It is noteworthy that the average variance retained for the first component was $87\%$ for the 100 simulations performed at time 7, $83\%$ at time 8, and $79\%$ at time 9. Then, only the first component is used for the analysis, so the model to estimate is:
\begin{align*}
\ln{\left(\dfrac{\pi_{it(miss)}}{1-\pi_{it(miss)}}\right)}= & \gamma_{0t}+\gamma_{1t}T_{it}+ \gamma_{2t}C_{it}\\
   \ln(\lambda_{it(miss)})= & \beta_{0t}+\beta_{1t}T_{it}+\beta_{2t}C_{it}, \; t=7,8,9\\
\end{align*}
where $\pi_{it(miss)}$ is the probability that the missing data is from the inflated zero model at time $t$, $\lambda_{it(miss)}$ is the mean of the Poisson model, $T_{it} $ is the treatment of the $i$-th plot and $C_{it}$ is the first eigenvalue of the principal component analysis at time $t$.
\end{enumerate}
\paragraph{Step 2.} With the estimated parameters obtained in Step 1, the following rule is used for the imputation of the lost information:
    \begin{equation*}
     \widehat{y}_{i1(miss)}=\left\{
         \begin{aligned}
                 & 0 && \text{if } \widehat{\pi}_{i1(miss)} \geq 0.5 ,\\
                 & [\widehat{\lambda}_{i1(miss)}] && \text{if } \widehat{\pi}_{i1(miss)}<0.5 ,\\
                 \end{aligned}\right.
\end{equation*}
The probability that a missing data is from the zero-inflated model is $\widehat{\pi}_{i1(miss)}$, it is proposed that if the value is greater than or equal to 0.5, the missing value is imputed as zero. Now, if $\widehat{\pi}_{i1(miss)}<0.5$ the value must belong to the Poisson model, and the imputed value is proposed to be $[\widehat{\lambda}_{i1(miss)} ]$ which is the function integer part of the mean of the Poisson model.  The values of $\widehat{\pi}_{1i(miss)}$ and $\widehat{\lambda}_{i1(miss)}$ are shown in tables \ref{p1pi} and \ref{p1lambda}, respectively.
\begin{table}[ht]
\centering
\begin{tabular}{rrrrrrrrrrr}
   \hline
  & We1 & We2 & We3 & We4 & We5 & We6 & We7 & We8 & We9 & Treat \\
   \hline
1 & 0 & 0 & 0 & 0 &\textbf{1.00}& 0 &\textbf{0.01}& 0 & 0 & 1 \\ 
  2 & 0 & 0 &\textbf{1.00 }& 0 & 0 & 0 & 0 &\textbf{0.56}& 0 & 1 \\ 
  3 & 0 &\textbf{1.00 }& 0 & 0 & 0 & 0 & 0 & 0 & 0 & 1 \\ 
  4 &\textbf{1.00 }& 0 &\textbf{1.00 }&\textbf{1.00 }& 0 &\textbf{0.02}& 0 & 0 & 1 & 1 \\ 
  5 &\textbf{1.00 }&\textbf{1.00 }& 0 & 0 &\textbf{0.00}& 0 & 0 & 0 & 1 & 1 \\ 
  6 &\textbf{1.00 }& 0 & 0 & 0 &\textbf{1.00}& 0 & 0 &\textbf{0.56}& 1 & 1 \\ 
  7 & 0 & 0 & 0 & 0 & 0 & 1 &\textbf{0.98}& 1 & 2 & 1 \\ 
  8 & 0 & 0 & 0 & 0 & 0 &\textbf{0.02}& 0 & 1 & 3 & 1 \\ 
  9 & 0 & 0 &\textbf{1.00 }& 0 & 0 & 0 & 0 & 0 & 0 & 2 \\ 
  10 & 0 & 0 & 0 & 0 & 0 & 0 & 0 & 0 & 0 & 2 \\ 
  11 & 0 & 0 & 0 & 0 & 0 & 0 & 0 & 0 & 0 & 2 \\ 
  12 & 0 & 0 & 0 & 0 & 0 & 0 & 0 & 1 & 0 & 2 \\ 
  13 & 0 & 0 & 0 & 1 & 1 & 1 & 0 & 1 & 0 & 2 \\ 
  14 & 0 & 0 & 0 & 0 & 1 &\textbf{0.00 }& 1 & 1 & 0 & 2 \\ 
  15 &\textbf{1.00 }& 0 &\textbf{1.00 }& 0 & 1 & 2 &\textbf{0.00 }&\textbf{0.00 }& 1 & 2 \\ 
  16 & 0 & 0 & 0 & 0 & 2 & 4 &\textbf{0.00 }& 2 & 3 & 2 \\ 
  17 & 0 & 0 & 0 &\textbf{1.00 }& 1 & 4 & 3 &\textbf{0.00 }& 2 & 3 \\ 
  18 & 0 & 0 & 0 & 0 & 1 & 5 & 4 &\textbf{0.00 }& 3 & 3 \\ 
  19 & 0 & 0 & 0 & 0 & 0 & 5 & 4 & 2 & 3 & 3 \\ 
  20 & 0 & 0 & 0 &\textbf{1.00 }& 0 & 5 & 5 & 2 & 4 & 3 \\ 
  21 & 0 & 0 & 0 & 0 & 0 & 4 &\textbf{0.00 }& 3 & 4 & 3 \\ 
  22 & 0 & 0 & 0 & 0 & 0 & 8 & 6 & 3 & 6 & 3 \\ 
  23 &\textbf{1.00 }& 0 &\textbf{1.00 }& 0 & 0 &\textbf{0.00 }& 7 & 4 & 4 & 3 \\ 
  24 &\textbf{1.00 }&\textbf{1.00 }& 0 & 0 & 0 & 9 &\textbf{0.98}&\textbf{0.00 }& 4 & 3 \\ 
    \hline
\end{tabular}
\caption{Weigths at each time $t$ for $\hat{\pi}_{it(miss)}$}\label{p1pi}
\end{table}

\begin{table}[ht]
\centering
\begin{tabular}{rrrrrrrrrrr}
   \hline
  & We1 & We2 & We3 & We4 & We5 & We6 & We7 & We8 & We9 & Treat \\
   \hline
1 & 0 & 0 & 0 & 0 &\textbf{0.00}& 0 &\textbf{0.00}& 0 & 0 & 1 \\ 
  2 & 0 & 0 &\textbf{0.00}& 0 & 0 & 0 & 0 &\textbf{0.62}& 0 & 1 \\ 
  3 & 0 &\textbf{0.00}& 0 & 0 & 0 & 0 & 0 & 0 & 0 & 1 \\ 
  4 &\textbf{0.00}& 0 &\textbf{0.00}&\textbf{0.00}& 0 &\textbf{0.17}& 0 & 0 & 1 & 1 \\ 
  5 &\textbf{0.00}&\textbf{0.00}& 0 & 0 &\textbf{0.00}& 0 & 0 & 0 & 1 & 1 \\ 
  6 &\textbf{0.00}& 0 & 0 & 0 &\textbf{0.00}& 0 & 0 &\textbf{0.62}& 1 & 1 \\ 
  7 & 0 & 0 & 0 & 0 & 0 & 1 &\textbf{0.00}& 1 & 2 & 1 \\ 
  8 & 0 & 0 & 0 & 0 & 0 &\textbf{0.17}& 0 & 1 & 3 & 1 \\ 
  9 & 0 & 0 &\textbf{0.00}& 0 & 0 & 0 & 0 & 0 & 0 & 2 \\ 
  10 & 0 & 0 & 0 & 0 & 0 & 0 & 0 & 0 & 0 & 2 \\ 
  11 & 0 & 0 & 0 & 0 & 0 & 0 & 0 & 0 & 0 & 2 \\ 
  12 & 0 & 0 & 0 & 0 & 0 & 0 & 0 & 1 & 0 & 2 \\ 
  13 & 0 & 0 & 0 & 1 & 1 & 1 & 0 & 1 & 0 & 2 \\ 
  14 & 0 & 0 & 0 & 0 & 1 &\textbf{3.21 }& 1 & 1 & 0 & 2 \\ 
  15 &\textbf{0.00}& 0 &\textbf{0.00}& 0 & 1 & 2 &\textbf{0.98 }&\textbf{1.02 }& 1 & 2 \\ 
  16 & 0 & 0 & 0 & 0 & 2 & 4 &\textbf{0.69 }& 2 & 3 & 2 \\ 
  17 & 0 & 0 & 0 &\textbf{0.00}& 1 & 4 & 3 &\textbf{2.67 }& 2 & 3 \\ 
  18 & 0 & 0 & 0 & 0 & 1 & 5 & 4 &\textbf{3.38 }& 3 & 3 \\ 
  19 & 0 & 0 & 0 & 0 & 0 & 5 & 4 & 2 & 3 & 3 \\ 
  20 & 0 & 0 & 0 &\textbf{0.00}& 0 & 5 & 5 & 2 & 4 & 3 \\ 
  21 & 0 & 0 & 0 & 0 & 0 & 4 &\textbf{4.86 }& 3 & 4 & 3 \\ 
  22 & 0 & 0 & 0 & 0 & 0 & 8 & 6 & 3 & 6 & 3 \\ 
  23 &\textbf{0.00}& 0 &\textbf{0.00}& 0 & 0 &\textbf{6.25 }& 7 & 4 & 4 & 3 \\ 
  24 &\textbf{0.00}&\textbf{0.00}& 0 & 0 & 0 & 9 &\textbf{6.71}&\textbf{1.40 }& 4 & 3 \\ 
    \hline
\end{tabular}
\caption{Weigths at each time $t$ for $\hat{\lambda}_{it(miss)}$}\label{p1lambda}
\end{table}

The parameters estimated for the above models are shown in Table \ref{table12345}.
{
\begin{table}[ht]
\centering
\begin{tabular}{rrrrrrrrrr}
   \hline
  & $t=1$ & $t=2$ & $t=3$ & $t=4$ & $t=5$ & $t=6$ &$t=7$ & $t=8$ & $t=9$  \\
   \hline
$\hat{\gamma}_{0t}$ & 25.30 &25.57  & 25.57   & 15.32 & 15.04  & -13.44 &-12.95 & -71.18 & -35.90 \\
   $\hat{\gamma}_{1t}$ & 0.00 &  0.00 & 0.00 & -25.87& -17.85 & 17.12 & 20.34 & -1.02 & 24.09 \\
   $\hat{\gamma}_{2t}$ &  0.00 &  0.00 & 0.00 & 0.00  &-24.43 & -9.49 & -92.53 & -1.02 & -8.02  \\
   $\hat{\gamma}_{3t}$ &  & &  &  &                   -19.85 & -13.87 & -99.19 & 48.61 & 23.17  \\
   $\hat{\beta}_{0t}$ & -25.30 &-25.30 & -25.30 &-21.30 & -19.28& -1.86& -18.83 & 0.19 & 0.61  \\
  $\hat{\beta}_{1t}$ & 0.00 &0.00 & 0.00       & -19.22 & 18.78 & 3.20 & 18.99& -0.298 & -0.44 \\
  $\hat{\beta}_{2t}$& 0.00 &0.00 &  0.00       &0.00   &  17.89 & 3.59& 20.12 &0.17 & 0.10\\
  $\hat{\beta}_{3t}$&  & & &  &                          0.00&  0.50& -0.55 &-0.46 & -0.28 \\
    \hline
\end{tabular}
\caption{Parameters estimated after using the proposed methodology}\label{table12345}
\end{table}
}
The estimated imputed values for all times are shown in Table \ref{p181}.
\begin{table}[ht]
\centering
\begin{tabular}{rrrrrrrrrrr}
   \hline
& We1 & We2 & We3 & We4 & We5 & We6 & We7 & We8 & We9 & Treat \\
   \hline
1 & 0 & 0 & 0 & 0 &\textbf{0.00}& 0 &\textbf{0.00}& 0 & 0 & 1 \\ 
  2 & 0 & 0 &\textbf{0.00}& 0 & 0 & 0 & 0 &\textbf{0.00}& 0 & 1 \\ 
  3 & 0 &\textbf{0.00}& 0 & 0 & 0 & 0 & 0 & 0 & 0 & 1 \\ 
  4 &\textbf{0.00}& 0 &\textbf{0.00}&\textbf{0.00}& 0 &\textbf{0.00}& 0 & 0 & 1 & 1 \\ 
  5 &\textbf{0.00}&\textbf{0.00}& 0 & 0 &\textbf{0.00}& 0 & 0 & 0 & 1 & 1 \\ 
  6 &\textbf{0.00}& 0 & 0 & 0 &\textbf{0.00}& 0 & 0 &\textbf{0.00}& 1 & 1 \\ 
  7 & 0 & 0 & 0 & 0 & 0 & 1 &\textbf{0.00}& 1 & 2 & 1 \\ 
  8 & 0 & 0 & 0 & 0 & 0 &\textbf{0.00}& 0 & 1 & 3 & 1 \\ 
  9 & 0 & 0 &\textbf{0.00}& 0 & 0 & 0 & 0 & 0 & 0 & 2 \\ 
  10 & 0 & 0 & 0 & 0 & 0 & 0 & 0 & 0 & 0 & 2 \\ 
  11 & 0 & 0 & 0 & 0 & 0 & 0 & 0 & 0 & 0 & 2 \\ 
  12 & 0 & 0 & 0 & 0 & 0 & 0 & 0 & 1 & 0 & 2 \\ 
  13 & 0 & 0 & 0 & 1 & 1 & 1 & 0 & 1 & 0 & 2 \\ 
  14 & 0 & 0 & 0 & 0 & 1 &\textbf{3 }& 1 & 1 & 0 & 2 \\ 
  15 &\textbf{0.00}& 0 &\textbf{0.00}& 0 & 1 & 2 &\textbf{1 }&\textbf{1 }& 1 & 2 \\ 
  16 & 0 & 0 & 0 & 0 & 2 & 4 &\textbf{1 }& 2 & 3 & 2 \\ 
  17 & 0 & 0 & 0 &\textbf{0.00}& 1 & 4 & 3 &\textbf{3 }& 2 & 3 \\ 
  18 & 0 & 0 & 0 & 0 & 1 & 5 & 4 &\textbf{3 }& 3 & 3 \\ 
  19 & 0 & 0 & 0 & 0 & 0 & 5 & 4 & 2 & 3 & 3 \\ 
  20 & 0 & 0 & 0 &\textbf{0.00}& 0 & 5 & 5 & 2 & 4 & 3 \\ 
  21 & 0 & 0 & 0 & 0 & 0 & 4 &\textbf{5 }& 3 & 4 & 3 \\ 
  22 & 0 & 0 & 0 & 0 & 0 & 8 & 6 & 3 & 6 & 3 \\ 
  23 &\textbf{0.00}& 0 &\textbf{0.00}& 0 & 0 &\textbf{6 }& 7 & 4 & 4 & 3 \\ 
  24 &\textbf{0.00}&\textbf{0.00}& 0 & 0 & 0 & 9 &\textbf{0.00}&\textbf{1 }& 4 & 3 \\ 
    \hline
\end{tabular}
\caption{Data imputation for all times.}\label{p181}
\end{table}
Comparing Table \ref{p181} with the original data Table \ref{origin1} shows that there are 36 matching observations for the success of the algorithm in this example of $79.5\%$.

\end{document}